\documentclass[12pt]{article}

\usepackage{epsf,amsfonts,hyperref}
\usepackage{cite}
\newcommand{\newsection}{    
\setcounter{equation}{0}
\section}
\renewcommand{\appendix}[1]{
    \addtocounter{section}{1} 
    \setcounter{equation}{0}
    \renewcommand{\thesection}{\Alph{section}}
    \newsection*{Appendix \thesection\protect\indent #1}
    \addcontentsline{toc}{section}{Appendix \thesection\ \ \ #1}
}
\newcommand\encadremath[1]{\vbox{\hrule\hbox{\vrule\kern8pt
\vbox{\kern8pt \hbox{$\displaystyle #1$}\kern8pt}
\kern8pt\vrule}\hrule}}
\def\enca#1{\vbox{\hrule\hbox{
\vrule\kern8pt\vbox{\kern8pt \hbox{$\displaystyle #1$}
\kern8pt} \kern8pt\vrule}\hrule}}

\newcommand\figureframex[3]{
\begin{figure}[bth]
\hrule\hbox{\vrule\kern8pt
\vbox{\kern8pt \vbox{
\begin{center}
{\mbox{\epsfxsize=#1.truecm\epsfbox{#2}}}
\end{center}
\caption{#3}
}\kern8pt}
\kern8pt\vrule}\hrule
\end{figure}
}
\newcommand\figureframey[3]{
\begin{figure}[bth]
\hrule\hbox{\vrule\kern8pt
\vbox{\kern8pt \vbox{
\begin{center}
{\mbox{\epsfysize=#1.truecm\epsfbox{#2}}}
\end{center}
\caption{#3}
}\kern8pt}
\kern8pt\vrule}\hrule
\end{figure}
}

\newcommand{\eq}[1]{eq.~(\ref{#1})}

\newcommand{\beq}{\begin{equation}}
\newcommand{\eeq}{\end{equation}}
\newcommand{\bea}{\begin{eqnarray}}
\newcommand{\eea}{\end{eqnarray}}

\renewcommand{\and}{{\qquad {\rm and} \qquad}}

\newcommand{\virg}{{\qquad , \qquad}}

\newcommand{\tr}{{\,\rm tr}\:}

\newcommand{\td}[1]{{\tilde{#1}}}

\renewcommand{\l}{\lambda}

\newcommand{\ee}[1]{{{\rm e}^{#1}}}

\renewcommand{\d}{{{\partial}}}
\newcommand{\D}{{{\hbox{d}}}}

\newcommand{\Pint}{{\int\kern -1.em -\kern-.25em}}

\newcommand\xx {{\cal X}}
\newcommand\yy {{\cal Y}}
\renewcommand\ss {\sigma}
\renewcommand\tt {\td{\sigma}}

\begin{document}
\sloppy


\pagestyle{empty}
\begin{flushright}
\hfill SPT-02/128\\
\hfill CRM-2868\\
\end{flushright}

\addtolength{\baselineskip}{0.20\baselineskip}
\begin{center}
\vspace{26pt}
{\large \bf {Large $N$ expansion of the 2-matrix model}}
\newline
\vspace{26pt}

{\sl B.\ Eynard}\hspace*{0.05cm}\footnote{ E-mail: eynard@spht.saclay.cea.fr }\\
\vspace{6pt}
Service de Physique Th\'{e}orique de Saclay,\\
F-91191 Gif-sur-Yvette Cedex, France.\\
\end{center}

\vspace{20pt}
\begin{center}
{\bf Abstract}
\end{center}
%

\begin{center}
We present a method, based on loop equations, to compute recursively,
 all the terms in the large $N$ topological expansion of the free energy for the 2-hermitian matrix model.
We illustrate the method by computing the first subleading term, i.e. the free energy of a statistical physics model on a discretized torus.
\end{center}



\newpage
\pagestyle{plain}
\setcounter{page}{1}


\newsection{Introduction}

Random matrix models \cite{Mehta, courseynard, ZJDFG, Guhr, BI, Moerbeke:2000} have a wide range of applications in mathematics and physics where they constitute a major field of activity.
They are involved in condensed matter physics (quantum chaos \cite{Guhr, QChaosHouches}, localization, crystal growths \cite{spohn},...etc),
 statistical physics \cite{ZJDFG, Matrixsurf, BIPZ, grossGQ2D} (on a 2d fluctuating surface, also called 2d euclidean quantum gravity, linked to conformal field theory),
 high energy physics (string theory \cite{DVV}, quantum gravity \cite{ZJDFG, grossGQ2D, GinspargGQ2D}, QCD \cite{Verbaarshot},...),
 and they are very important in mathematics too: (they seem to be linked to the Riemann conjecture \cite{Mehta, Odl}), they are important in combinatorics,
 and provide a wide class of integrable systems \cite{BI, HTW, McLaughlin}.

In the 80's, random matrix models were introduced as a toy model of 0-dimensional string theory and quantum gravity \cite{Matrixsurf, ZJDFG, BIPZ}.
By fine-tuning the parameters of the model, one can reach many multi-critical points, which are in relationship with the $(p,q)$ minimal conformal models \cite{Matrixtwoconform, DKK}.
The 1-hermitian matrix models is limited to $(p,2)$ conformal minimal models,
 whereas the 2-matrix model can represent any $(p,q)$ conformal minimal model \cite{DKK}.
The 2-matrix model is thus more general than the 1-matrix model,
 and it has been the source of considerable interest in the past few months \cite{AvM2, McLaughlin, BEHduality, BEHRH, BEHneeds, BEHansatz, BEHAMS, guionnet, PZJZ}.

The free energy of the 1-matrix model was conjectured \cite{thoft, ZJDFG, ACM, ACKM}
and also rigorously proven \cite{EML} to have a $1/N^2$ expansion
 called topological expansion ($N$ is the size of the matrix):
\beq
F = \sum_{h=0}^\infty N^{-2h} F^{(h)}
\eeq
The authors of \cite{ACM} invented an efficient method to compute recursively all the $F^{(h)}$'s, and they improved it in \cite{ACKM}.

The 2-matrix model is conjectured to have a similar $1/N^2$ topological expansion.
Indeed, this expansion is the main motivation for applications to 2-dimensional quantum gravity \cite{ZJDFG, courseynard},
 because each $F^{(h)}$ is the partition function of a statistical physics model on a genus $h$ surface.
There is at the present time no rigorous proof\footnote{The Riemann-Hilbert approach seems to be the best way to prove the existence of the $1/N^2$ topological extension as in \cite{EML}. The Riemann-Hilbert problem for the 2-matrix model has been formulated \cite{BEHRH, BEHAMS, BEHansatz, Kapaev}, and seems to be on the verge of being solved \cite{BEHansatz}.} of the existence of such an expansion,
but, assuming that it exists, the aim of the present work is to give a method to compute recursively the terms of the expansion,
 similar to that of \cite{ACM}.

\bigskip

The 2-matrix model was first introduced as a model for two-dimensional gravity, with matter, and in particular with an Ising field \cite{Kazakov, KazakovIsing}.
The diagrammatic expansion of the 2-matrix model's partition function is known to generate 2-dimensional statistical physics models on a random discrete surface \cite{ZJDFG, Matrixsurf, Kazakov}:
\beq
N^2 F=-\ln{Z} = \sum_{\rm surfaces}\,\,  \sum_{\rm matter}  \ee{-{\rm Action}}
\eeq
where the Action is the matter action (like Ising's nearest neighboor spin coupling) plus the gravity action (total curvature and cosmological constant) \cite{ZJDFG}.
The cosmological constant couples to the area of the surface, and $N$ (the size of the matrix) couples to the total curvature, i.e. the genus of the surface.
The large $N$ expansion thus generates a genus expansion:
\beq
F=\sum_{h=0}^\infty N^{-2h} F^{(h)}
\eeq
where $F^{(h)}$ is the partitrion function of the statistical physics model on a random surface of fixed genus $h$.
\beq\label{Fhsurfaces}
F^{(h)} = \sum_{{\rm genus}\, h\, {\rm surfaces}}\,\,  \sum_{\rm matter}  \ee{-{\rm Action}}
\eeq
The leading term $F^{(0)}$ is the planar contribution.
It can be computed by many different methods,
for instance the saddle point method (along the method invented by \cite{matytsin} and rigorously established by \cite{guionnet})
or the loop equation method which we will explain below.
Our goal in this article is to compute $F^{(1)}$
and present an algorithmic method for computing $F^{(h)}$ for $h\geq 1$.
We generalize the method of \cite{ACM}.

\subsection{Reminder: the one-matrix model}

For a given polynomial $V(x)$ of degree $d+1$, called the potential, and a given integer $N$,
we define the partition function $Z$ and the free energy $F$ as:
\beq
Z = \ee{-N^2 F} = \int \D{M} \ee{-N\tr V(M)}
\eeq
where the integral is over the set of hermitian matrices of size $N$, with the measure $\D{M}$ equal to the product of Lebesgue measures of all real components of $M$.

The free energy has a $1/N^2$ power series expansion (under the 1-cut asumption, this will be discussed in more details in section \ref{cutgenusasumption}):
\beq
F = \sum_{h=0}^\infty N^{-2h} F^{(h)}
\eeq

In their pioneering work, the authors of \cite{ACM} invented a method to compute recursivelly all the $F^{(h)}$'s.
In particular they found that the genus one free energy for the 1-matrix model is:
\beq\label{F11mat}
F^{(1)} = -{1\over 24} \ln{\left((b-a)^4M(a)M(b)\right)}
\eeq
where $b,a$ and the polynomial $M(x)$ are related to the large $N$ limit of the average density of eigenvalues of the matrix:
\beq
\rho(x) = {1\over 2\pi} M(x) \sqrt{(x-a)(b-x)}
\virg
M(x) = \mathop{\rm Pol}_{x\to\infty}\,\, V'(x)/\sqrt{(x-a)(x-b)}
\eeq
This density has a compact support $[a,b]$, and $M(x)$ is a polynomial of degree $d-1$ (the 1-matrix case is re-explained in section \ref{section1mat}).

We are going to extend this kind of expression (\eq{F11mat}) for the 2-matrix model.

\subsection{Outline cof the article}

\begin{itemize}
\item In section \ref{section2matintro} we introduce the definitions and notations, in particular we define the 1-loop functions and 2-loop functions, and the loop-insertion operators.

\item In section \ref{sectionloopequations} we explain the loop equation method, and derive the ``Master loop equation''.

\item In section \ref{sectiontheloopequation}, we observe that, to leading order, the master loop equation is an algebraic equation of genus zero,
and we study the geometry and the sheet-structure of the underlying algebraic curve.

\item In section \ref{sectionleadingorder} we use the algebraic master loop equation to compute all the previously defined loop functions to leading order.

\item In section \ref{sectionsubleading} we include the previously neglected $1/N^2$ term in the loop equation,
and we compute the 1-loop function $Y(x)$ to next to leading order.
Then we derive the next to leading order free energy $F^{(1)}$ by integrating $Y^{(1)}(x)$.
We also discuss how to compute higher order terms.

\item In section \ref{section1mat} we compare with the results previously known for the one matrix case.

\item section \ref{conclusion} is the conclusion.

\end{itemize}

\newsection{The 2-matrix model}
\label{section2matintro}

\subsection{The partition function}

For two given polynomials $V_1(x)$ of degree $d_1+1$ and $V_2(y)$ of degree $d_2+1$, called the potentials, and a given integer $N$,
we define the partition function $Z$ and the free energy $F$ as:
\beq\label{Zdef}
Z= \ee{-N^2 F} = \int \D{M_1} \D{M_2}\, \ee{-N\tr [ V_1(M_1)+V_2(M_2)-M_1 M_2 ]}
\eeq
where the integral is over the set of pairs of hermitian matrices $M_1$ and $M_2$ of size $N$, with the measure $\D{M_1}\D{M_2}$ equal to the product of Lebesgue measures of all real components of $M_1$ and $M_2$ (see \cite{ZJDFG} for definitions).

The potentials $V_1$ and $V_2$ are written:
\beq\label{Vdef}
V_1(x) = \sum_{k=0}^{d_1} {g_{k+1}\over k+1} x^{k+1}
\, , \quad
V_2(y) = \sum_{k=0}^{d_2} {g^*_{k+1}\over k+1} y^{k+1}
\eeq

\medskip
{\noindent \bf Remark:} \eq{Zdef} is well defined only when the potentials are bounded from below,
i.e. when:
\beq\label{Zexistencecondition}
\left\{ d_1 \,\,\, {\rm and} \,\, d_2 \,\,\, {\rm are \,\, odd} \qquad g_{d_1+1}>0\, ,  \,\, g^*_{d_2+1}>0 \right\}
\eeq
However, it is known \cite{ZJDFG} how to give a meaning to \ref{Zdef} for potentials non bounded from below, in the large $N$ limit
(somehow the tunnel effect allowing the matrices to ``escape from the potential wells'' is suppressed like $O(\ee{-N})$ in the large $N$ limit, it plays no role in the $1/N^2$ expansion).
Therefore, we will consider arbitrary polynomial potentials which do not necessarily satisfy condition (\ref{Zexistencecondition}).
For simplicity, we will assume that the potentials $V_1$ and $V_2$ are real\footnote{The derivation of the loop equations in section \ref{sectionloopequations} should be slightly modified in the case of non-real potentials, because some change of variables we consider would be no longer hermitian.
But most of the results derived in this article are still valid for complex potentials.}.

In addition, we will assume that the potentials $V_1$ and $V_2$ are generic, i.e. they are not critical (we will explain what is a critical potential in section \ref{sectionendpoints}, see \cite{DKK}).

\medskip
The partition function $F$ can be interpreted (through the Feynmann graphs expansion, see \cite{ZJDFG, courseynard})
 as the generating function of surfaces made of polygons with a spin:
spin ``up'' (+) for the polygons generated by the vertices of $M_1$, ``down'' (-) for the polygons generated by the vertices of $M_2$.
\beq\label{Fsurfaces}
F = \sum_{S\in {\cal E}} N^{-2h(S)}\,  {1\over \#\,{\rm Aut}(S)} \, {1\over g_2 g_2^*}\, \prod_{k=1, k\neq 2}^{d_1+1} (-g_k)^{n_k(S)} \, \prod_{j=1, j\neq 2}^{d_2+1} (-{g^*}_j)^{\td{n}_j(S)}
\eeq
where ${\cal E}$ is the ensemble of surfaces made of up-polygons
 with at most $d_1+1$ sides and down-polygons with at most $d_2+1$ sides.
Aut$(S)$ is the group of automorphisms of the graph of edges of $S$,
 $h(S)$ is the genus of $S$,
 $n_k(S)$ ($1\leq k \leq d_1+1$) is the number of up $k$-gons of $S$ and $\td{n}_j(S)$ ($1\leq j \leq d_2+1$) is the number of down $j$-gons of $S$.\par
We write digrammatically:
$$ F = {\epsfysize=3.truecm\epsfbox{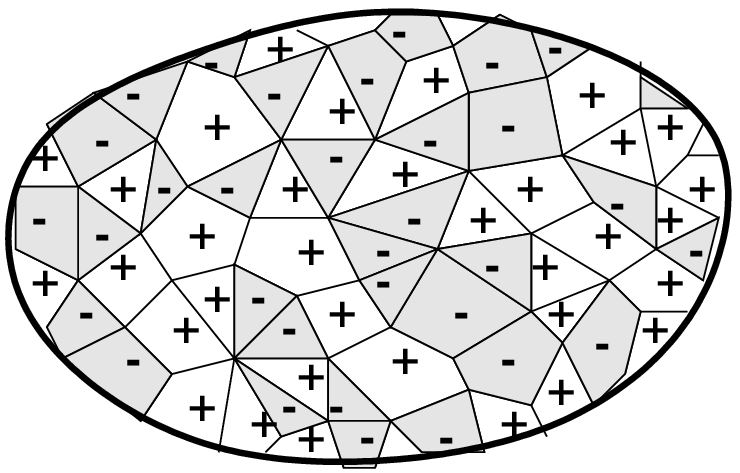}} + {1\over N^2} {\epsfysize=3.truecm\epsfbox{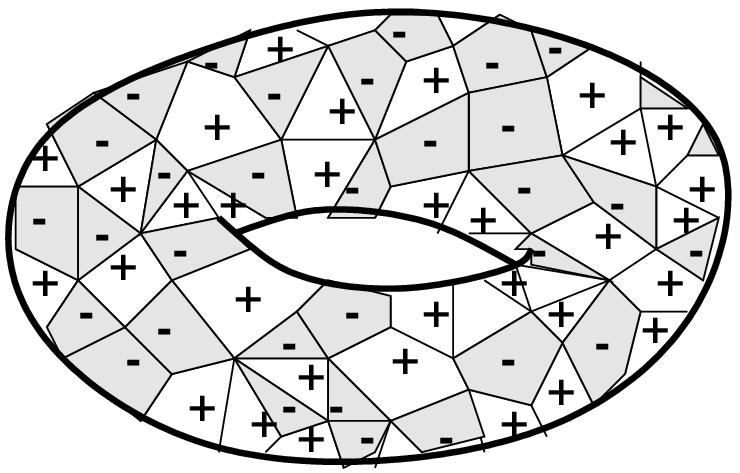}} + \dots $$

This is why the two-matrix model can represent the Ising model on a random discretized surface \cite{Kazakov, KazakovIsing}, and other 2d statistical physics models on a random surface carrying some type of matter \cite{ZJDFG}.
It is clear from \eq{Fsurfaces} that relevant applications to 2d statistical physics on a random surface involve potentials which violate condition \ref{Zexistencecondition} (all the $g_k$'s and $\td{g}_l$'s must be negative in order to have positive weights).

\subsection{Definition of the loop functions}

We introduce the following expectation values computed with the probability weight ${1\over Z} \ee{-N\tr [V_1(M_1)+V_2(M_2)-M_1 M_2]}$:
\beq
T_{k,l} = {1\over N} \left< \tr M_1^k M_2^l \right>
\eeq
We will be particularly interested in their large $N$ limits and large $N$ expansion.

$T_{k,l}$ can be interpreted as the partition function of a statistical physics model
on surfaces with one boundary of lenght $k+l$ ($k$ contiguous + and $j$ contiguous -):
$$ T_{k,l} = {\epsfysize=3.truecm\epsfbox{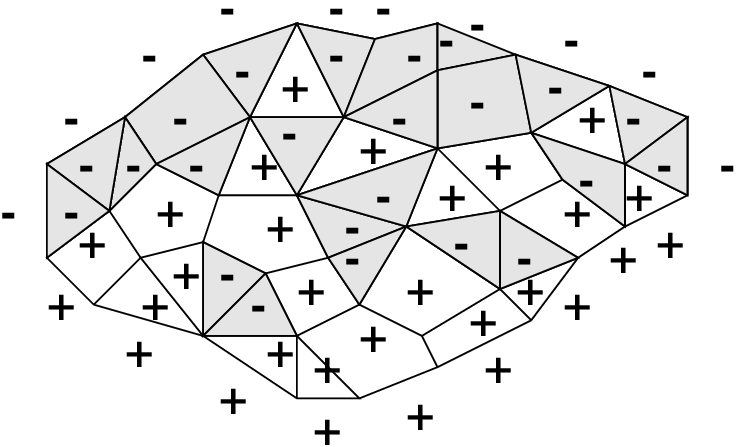}} + {1\over N^2}{\epsfysize=3.truecm\epsfbox{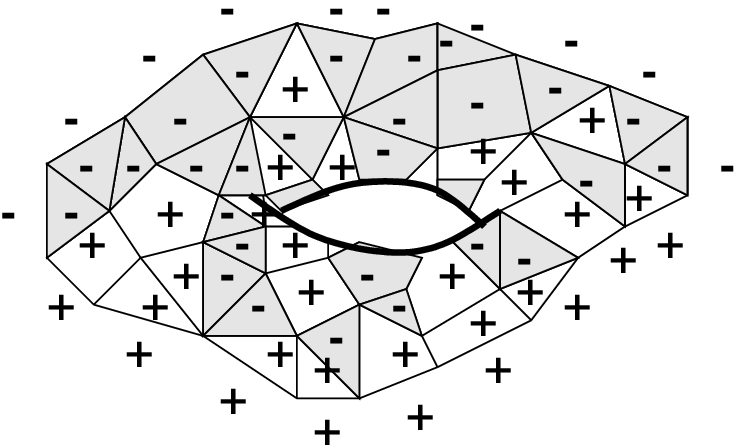}} + \dots $$

It is more convenient to introduce generating functions which contain all the $T_{kl}$'s at once:

\subsubsection{One-loop functions}

We define the following formal functions of the complex variables $x$ and $y$:
\beq\label{defWx}
W(x) = {1\over N} \left< \tr {1\over x-M_1} \right> = \sum_{k=0}^\infty {T_{k,0}\over x^{k+1}} \virg Y(x) = V'_1(x)-W(x)
\eeq
\beq\label{deftdWy}
\td{W}(y) = {1\over N} \left< \tr {1\over y-M_2} \right> = \sum_{k=0}^\infty {T_{0,k}\over y^{k+1}} \virg X(y) = V'_2(y)-\td{W}(y)
\eeq
These functions are formally defined only through their large $x$ and large $y$ expansions,
they are only a convenient rewritting of the collection of $T_{k,0}$ and $T_{0,k}$.
The sum is not necessarily convergent.

We will find below, that $W^{(0)}(x) = {\rm lim}_{N\to\infty} W(x)$ (resp. $\td{W}(y)$) satisfies an algebraic equation,
 and thus $W^{(0)}(x)$ (resp. $\td{W}^{(0)}(y)$) is an analytical function of $x$ (resp. $y$), in the complex plane, with a cut.
The radius of convergence of \eq{defWx} is finite in the large $N$ limit.
Their analytical structure will be dicussed in details in part \ref{sectionsheetstructure}.

The functions $W(x)$ and $\td{W}(y)$ are called the resolvents,
 they provide most of the information about the statistical properties of the spectra of $M_1$ and $M_2$.
For instance the location of the cut of $W^{(0)}(x)$,
is the support of the large $N$ average density of eigenvalues of $M_1$,
and in the large $N$ limit, the average density of eigenvalues is proportional to the discontinuity of $W^{(0)}(x)$ along its cuts.

Their diagrammatic expansion is the generating function for statistical physics models coupled to gravity,
 on a surface with one boundary (a disc if genus zero),
 this is why they are called one-loop functions.
\beq
W(x) = \sum_{l_1=0}^\infty x^{-l_1-1} \,\,\, \sum_{{\hbox{\tiny  surfaces, }} \atop {\hbox{\tiny boundary $=l_1$ up}} }\,\,  \sum_{\rm matter}  \ee{-{\rm Action}}
= {\epsfysize=2.truecm\epsfbox{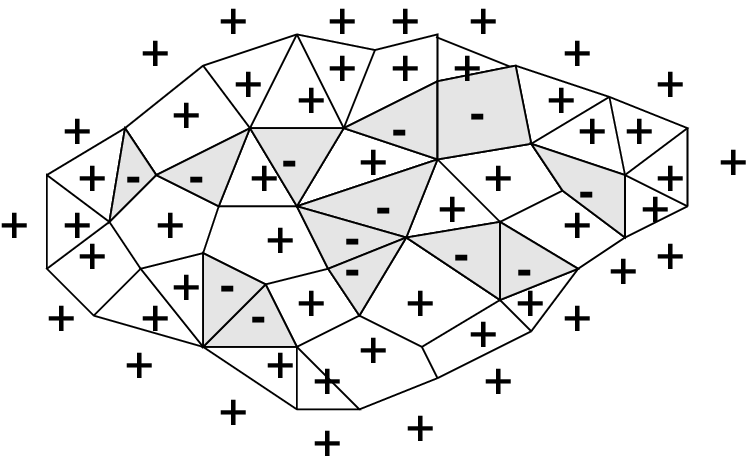}} + \dots
\eeq
where the sum carries on all discretized surfaces with
 one boundary made of $l_1$ spins +.
\beq
\td{W}(y) = \sum_{l_2=0}^\infty x^{-l_2-1} \,\,\, \sum_{{\hbox{\tiny  surfaces, }} \atop {\hbox{\tiny boundary $=l_2$ down}} }\,\,  \sum_{\rm matter}  \ee{-{\rm Action}}
= {\epsfysize=2.truecm\epsfbox{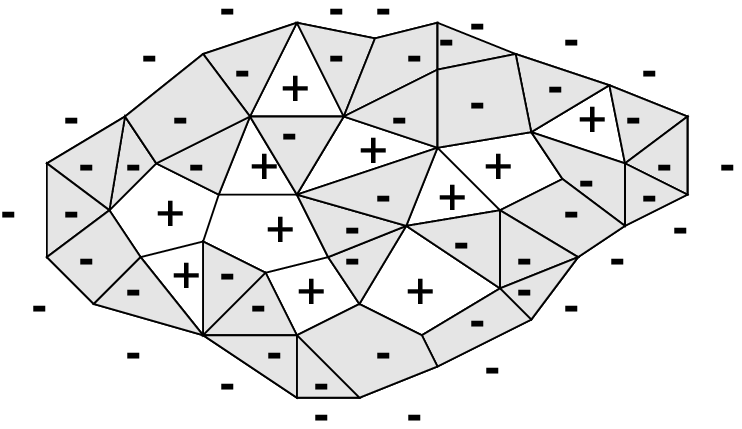}} + \dots
\eeq
where the sum carries on all discretized surfaces with
 one boundary made of $l_2$ spin -.

We also define:
\beq
W(x,y) = {1\over N} \left< \tr {1\over x-M_1} {1\over y-M_2} \right>
= \sum_{k=0}^\infty \sum_{l=0}^\infty {T_{k,l}\over x^{k+1} y^{l+1}}
\eeq
This function provides information on the correlation between the spectrum of $M_1$ and that of $M_2$.
Its diagrammatic expansion generates surfaces with one ``bi-colored'' boundary, i.e. a boundary of lenght $l_1+l_2$ made of $l_1$ consecutive spins up, followed by $l_2$ down:
\beq
W(x,y) = \sum_{l_1=0}^\infty \sum_{l_2=0}^\infty x^{-l_1-1}\, y^{-l_2-1} \,\,\, \sum_{{\hbox{\tiny  surfaces, }}\atop {\hbox{\tiny bnd $=l_1$ up + $l_2$ down}} }\,\,  \sum_{\rm matter}  \ee{-{\rm Action}}
= {\epsfysize=2.truecm\epsfbox{Wxy0.eps}}
\eeq

We also define:
\beq
U(x,y) = {1\over N} \left< \tr {1\over x-M_1} {V'_2(y)-V'_2(M_2)\over y-M_2} \right>
\eeq
\beq
\td{U}(x,y) = {1\over N} \left< \tr {V'_1(x)-V'_1(M_1)\over x-M_1} {1\over y-M_2} \right>
\eeq
which are polynomials in one of the variable and analytical near $\infty$ for the other variable.

And we define
\beq
P(x,y)  = {1\over N} \left< \tr  {V'_1(x)-V'_1(M_1)\over x-M_1} {V'_2(y)-V'_2(M_2)\over y-M_2} \right>
\eeq
which is a polynomial in both variables, of degree $(d_1-1,d_2-1)$.

\bigskip
In appendix C we compute another one-loop function $H(x,y,x',y')$,
 which generates surfaces with one quadri-colored boundary,
 i.e. surfaces whose boundary is +-+-.
Somehow it is the random surface analogus of a square with opposite sides of the same spin and adjacent sides of opposite spin.
It maybe very usefull to compute that kind of function to study how spin percolates from one side to the other.
It may have applications to compute boundary operators in 2d statistical physics on a random surface. 

\subsubsection{Two-loop functions}

We also need to define the following correlation functions, which contain the product of two traces.
Their diagrammatic expansion is the generating function for statistical physics models coupled to gravity, on a surface with two boundaries (cylinder-like),
 this is why they are called two-loop functions.

\beq
\Omega(x;x') = {\d W(x)\over \d V_1(x')} = \left< \tr {1\over x-M_1}  \tr {1\over x'-M_1}\right> - N^2 W(x) W(x')
\,\,\, ,
\eeq

\beq
\td\Omega(y;x') = {\d \td{W}(y) \over \d V_1(x')} = \left< \tr {1\over x'-M_1}  \tr {1\over y-M_2}\right> - N^2 W(x') \td{W}(y)
\,\,\, ,
\eeq

 \beq
U(x,y;x') = {\d U(x,y) \over \d V_1(x')} = \left< \tr {1\over x-M_1} {V'_2(y)-V'_2(M_2)\over y-M_2} \tr {1\over x'-M_1} \right> - N^2 U(x,y)W(x') \,\,\, ,
\eeq

\beq
\td{U}(x,y;x') = {\d \td{U}(x,y) \over \d V_1(x')} = \left< \tr {V'_1(x)-V'_1(M_1)\over x-M_1} {1\over y-M_2} \tr {1\over x'-M_1} \right> - N^2 \td{U}(x,y)W(x')
\,\,\, ,
\eeq

\bea
P(x,y;x') = {\d P(x,y) \over \d V_1(x')} & = & \left< \tr  {V'_1(x)-V'_1(M_1)\over x-M_1} {V'_2(y)-V'_2(M_2)\over y-M_2} \tr {1\over x'-M_1} \right> \cr
&&  - N^2 P(x,y) W(x')
\,\,\, ,
\eea
where the ``loop-insertion'' operators (see \cite{ACM}) are
formally\footnote{By formally, we mean they make sense only through their
 large $x$ (resp. $y$) expansion.
 If $k$ is larger than $d_1$, $\d/\d g_k$ means:
 take a potential of degree $\geq k$,
 and then take the derivative w.r.t. $g_k$ at $g_j=0$ for $j>d_1$.}
 defined as:
\beq
{\d\over \d V_1(x')} = \sum_{k=1}^\infty {k\over x'^k} {\d\over \d g_k}
\virg
{\d\over \d V_2(y')} = \sum_{k=1}^\infty {k\over y'^k} {\d\over \d g^*_k}
\eeq
their action on the partition function or on expectation values inserts a trace,
i.e. a boundary (a ``loop'') for the ensemble of surfaces.
For instance we have:
\beq
W(x) = {\d\over \d V_1(x)} F \virg \Omega(x;x') = {\d\over \d V_1(x')}W(x) ={\d\over \d V_1(x)}{\d\over \d V_1(x')}F = \Omega(x';x)
\eeq
The diagrammatic expansion of $F$ gives surfaces with no boundary, $W(x)$ gives surfaces with one boundary,
 $\Omega(x,x')$ gives surfaces with two (spin up) boundaries,
and $\td\Omega(x,y)$ gives surfaces with one spin up boundary and one spin down boundary.
And so on: ${\d\over \d V_1(x')}$ inserts a + loop, ${\d\over \d V_2(y')}$ inserts a - loop.
$$ \Omega(x;x') = {\epsfysize=3.truecm\epsfbox{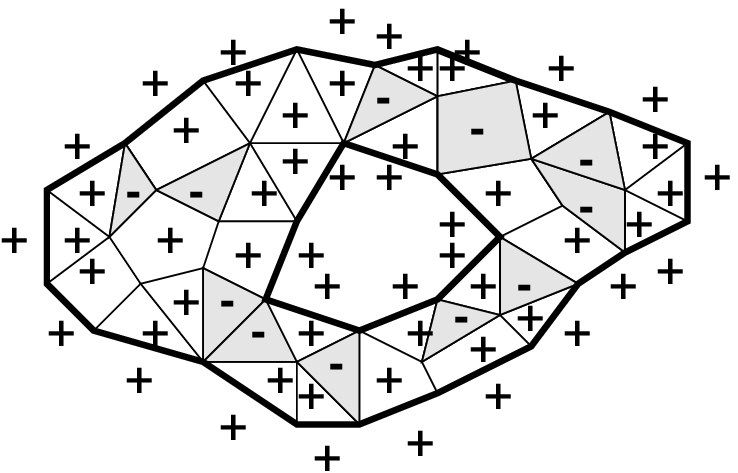}}
\virg
\td\Omega(y;x') = {\epsfysize=3.truecm\epsfbox{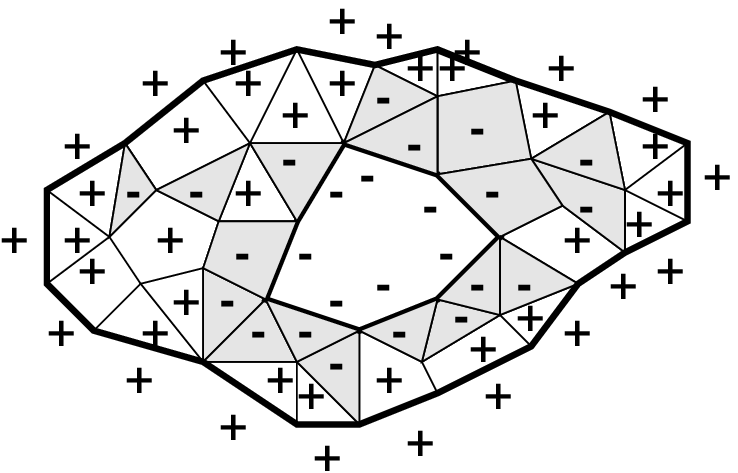}} \,\, .
$$

\newsection{The loop equations}
\label{sectionloopequations}

\subsection{Generalities}

The loop equations are a consequence of the reparametrization invariance of the matrix integral.
Basically, one writes that under an infinitesimal change of variable of the type:
\beq
M_1 \to \td{M}_1 = M_1 + \epsilon f(M_1,M_2)
\eeq
the partition function is left unchanged to order 1 in $\epsilon$.
We have to compute the variation of the measure and the variation of the potential:
\bea
Z & = &  \int \D{\td{M}_1} \D{M_2}\, \ee{-N\tr [ V_1(\td{M}_1)+V_2(M_2)-\td{M}_1 M_2 ]} \cr
& = & \int \D{M_1} \D{M_2} \, (1+\epsilon J(M_1,M_2)) \, \ee{-N\tr [ V_1(M_1)+V_2(M_2)-M_1 M_2 ]} (1-\epsilon K(M_1,M_2)) \cr
\eea
where $J(M_1,M_2)$ is the jacobian of the change of variable computed to order $1$ in $\epsilon$:
\beq
\det{\d \td{M_1}\over \d M_1} = 1+\epsilon J(M_1,M_2) + O(\epsilon^2)
\eeq
and $K(M_1,M_2)$ is the variation of the action to order $1$ in $\epsilon$.
We must have:
\beq
\left< J(M_1,M_2)\right> = \left< K(M_1,M_2)\right>
\eeq
We are going to give the receipe\footnote{Easy to prove directly order by order in $x$. We leave the proof to the reader. A similar method is used in \cite{BECK}.}
 to compute $J(M_1,M_2)$ for various change of variables $f(M_1,M_2)$.\par

\medskip

{\noindent \em Remarks}:\par
- It is important to take into account the non-commutativity of matrices.\par
- It is also important to choose a change of variable which is a \underline{hermitian} matrix.\par
- We are presenting the method for variations of $M_1$ but of course the same applies to $M_2$ and we will use both changes of variables in the body of this article.

\subsection{split and merge rules}

We will be interested in change of variables of the following
form\footnote{As before,
${1\over x-M_1}=\sum_{k} x^{-k-1} M_1^k$ is a ``formal'' notation for the collection of $M_1^k$, $k=0\dots\infty$.
}:

\begin{itemize}
\item  $f(M_1,M_2) = A{1\over x-M_1}B $:\par
where the matrices $A$ and  $B$ are functions of $M_1$ and $M_2$.
The variation of the measure is:
\beq
J(M_1,M_2) = \tr\left( A{1\over x-M_1}\right) \tr\left( {1\over x-M_1} B\right)  + {\rm contributions\, from\, }A,\, {\rm and}\, B.
\eeq
somehow, you split the trace into two traces whenever you meet a $1\over x-M_1$ term outside a trace, and each side receives a $1\over x-M_1$.\par

\item $f(M_1,M_2) = A \tr\left( B{1\over x-M_1} \right)$:\par
where $A$, $B$ are functions of $M_1$ and $M_2$.
The variation of the measure is:
\beq
J(M_1,M_2) = \tr\left( A{1\over x-M_1}B{1\over x-M_1}\right)  + {\rm contributions\, from\, }A,\, {\rm and}\, B.
\eeq
somehow, you merge the traces whenever you meet a $1\over x-M_1$ term inside a trace, and duplicate the $1\over x-M_1$ term (one before $B$, one after $B$).\par

\item The variations of $A$ and $B$ are computed recursively by the chain rule.
\end{itemize}

\medskip

The variation of the action is simply:
\beq
{1\over N}K(M_1,M_2) = \tr \left( V'_1(M_1) f(M_1,M_2) - M_2 f(M_1,M_2) \right)
\eeq
and we will use the following trick very often:
\beq
\tr\left( V'_1(M_1) {1\over x-M_1} A \right)= V'_1(x) \tr\left( {1\over x-M_1} A  \right) - \tr \left({V'_1(x)-V'_1(M_1)\over x-M_1} A\right)
\eeq
where it is important to notice that the second term is polynomial in $x$.

\subsection{Factorization ``theorem''}

Then, we will make the asumption that in the large $N$ limit, the average of product of traces decouple:
\beq
{1\over N^2}\left< \tr A \tr B \right> \sim {1\over N}\left< \tr A \right> {1\over N}\left< \tr B \right> + O(1/N^2)
\eeq
more precisely that the connected correlation function
\beq
\left< \tr A \tr B \right>_{\rm c} = \left< \tr A \tr B \right> -\left< \tr A \right> \left< \tr B \right>
\eeq
has a large $N$ limit.

The factorization ``theorem'' is not a theorem,
 it was never proven for the 2-matrix-model\footnote{It was proven rigorously for the one-matrix model in \cite{EML}.},
 but there are plenty of heuristic reasons to believe that it holds,
 and it was used by nearly all authors in this field.
As we said in the introduction,
 we are not going to prove the existence of the $1/N^2$ expansion,
 but assuming it, we are going to give an algorithmic method to compute it.

\subsection{The master loop equation}

Consider the change of variable:
$$ M_2 \to M_2 + \epsilon {1\over x-M_1}$$
which is indeed a hermitian matrix.
It leads to (there is no variation of the measure since $\delta M_2$ is independent of $M_2$, the RHS is the variation of the potential):
\beq
0 =  \left< \tr {1\over x-M_1}V'_2(M_2) - {M_1\over x-M_1} \right>
\eeq
i.e.
\beq\label{loopeqxV2}
{1\over N} \left< \tr {1\over x-M_1}V'_2(M_2) \right> = x W(x) - 1
\eeq

\medskip

Now, consider the change of variables (which is indeed a hermitian matrix):
$$ M_1 \to M_1 + \epsilon {V'_2(y)-V'_2(M_2)\over y-M_2} {1\over x-M_1} + \epsilon {1\over x-M_1} {V'_2(y)-V'_2(M_2)\over y-M_2} $$
it leads to (the LHS is the variation of the measure, the RHS of the potential, and we use the cyclicity of the trace):
\bea
&& \left< \tr {1\over x-M_1} \tr {1\over x-M_1}{V'_2(y)-V'_2(M_2)\over y-M_2}\right>
  = \\
&& N\left< \tr {V'_1(M_1)\over x-M_1}{V'_2(y)-V'_2(M_2)\over y-M_2}  -
 \tr{1\over x-M_1} M_2{V'_2(y)-V'_2(M_2)\over y-M_2}  \right>
\eea
i.e. (using \eq{loopeqxV2}):
\beq\label{eqUxyexact}
(y-Y(x)) U(x,y) = V'_2(y)W(x) - P(x,y) - x W(x) + 1 - {1\over N^2} U(x,y;x)
\eeq

Notice that $U(x,y)$ is a polynomial in $y$, therefore it is finite for $y=Y(x)$.
If we choose $y=Y(x)$ we get (using $W(x)=V'_1(x)-Y(x)$):
\beq
\left\{
\begin{array}{l}
y=Y(x) \cr
(V'_2(y)-x)(V'_1(x)-y) - P(x,y) + 1  = {1\over N^2} U(x,y;x)
\end{array}
\right.
\eeq

which we write:
\beq\label{loopequation}
\encadremath{
E(x,Y(x)) = {1\over N^2} U(x,Y(x);x)
}\eeq
where $E(x,y)$ is a polynomial in $x$ (degree $d_1+1$) and $y$ (degree $d_2+1$):
\beq
\encadremath{
E(x,y) = (V'_1(x)-y)(V'_2(y)-x) - P(x,y) +1
}\eeq
\eq{loopequation} is called the ``master loop equation'' \cite{staudacher},
we will see below that it allows to determine the function $Y(x)$ and all other functions in the problem.

\subsection{$1/N^2$ expansion}

We assume that the function $Y(x)$ can be written as a power series in $1/N^2$:
\beq
Y(x) = Y^{(0)}(x) + {1\over N^2} Y^{(1)}(x)  + \dots
\eeq
Such an expansion is believed to hold in the ``1-cut case'', which will be explained below (section \ref{cutgenusasumption}).

\begin{itemize}
\item
To leading order, the loop equation is an algebraic equation:
\beq
E^{(0)}(x,Y^{(0)}(x)) = 0
\eeq
where
\beq
E^{(0)}(x,y) = (V'_1(x)-y)(V'_2(y)-x)-P^{(0)}(x,y)+1
\eeq
The polynomial $P^{(0)}(x,y)$ has degrees $(d_1-1,d_2-1)$,
 it is determined by the large $x$ and large $y$ behaviors of $Y(x)$ and $X(y)$, as well as some analytical requirement, namely the ``one-cut asumption''.
As we will see below, in the 2-matrix model, this asumption is replaced by the asumption that the algebraic curve $E(x,y)=0$ has genus zero.
In the next section we will study this algebraic curve, and we will explain how to compute the polynomial $P^{(0)}(x,y)$.

Let us note that this algebraic equation was derived by amny authors and various methods in the literature:
withe the loop equation method, the existence of an algebraic equation is proven (and computed for small degree potentials) in \cite{staudacher}
, and was presented for arbitrary polynomial potentials in the appendix of \cite{eynard} and \cite{eynardchain}.
The authors of \cite{PZJ} and \cite{PZJZ} derived it by the ``saddle point method'' inspired from \cite{matytsin}, and from the 2-toda hierarchy dispersionless limit,
and that algebraic equation was used by many other authors in particular cases.
And notice that it is the same equation as the spectral curve of any of the 4 dual differential systems in \cite{BEHneeds}.

\item
To order $1/N^2$ we expand:
\beq
E(x,y) = E^{(0)}(x,y) - {1\over N^2} P^{(1)}(x,y) + \dots
\eeq
where $P^{(1)}(x,y)$ has degrees $(d_1-1,d_2-1)$.
We will see how to compute it in section \ref{sectionsubleading}.

We easily get:
\beq\label{loopeqY1}
Y^{(1)}(x) = {U^{(0)}(x,Y^{(0)}(x);x) + P^{(1)}(x,Y^{(0)}(x))\over \d_y E^{(0)}(x,Y^{(0)}(x)) }
\eeq
In this expression, all terms in the RHS are computed to leading order, except the polynomial $P^{(1)}(x,y)$.
But $P^{(1)}(x,y)$ is completely determined (in terms of leading order functions only) by the condition that
 all the unwanted zeroes of the denominator (those which don't correspond branch-points) cancel.
In other words, once we know the functions $Y(x)$ and $U(x,y;x')$ to leading order, we can compute $Y(x)$ to order~$1$.
Similarly, we can compute every loop function to order~$1$.

\item
We can repeat the procedure to find higher orders.
Once we know all the loop functions up to order $h$, we can compute them to order $h+1$.
Of course the expressions become more and more complicated with increasing orders, but the method is the same.
We will illustrate our method by computing the order $1$ only.

\item once we know $Y^{(h)}(x)$, we can find the order $h$ free energy $F^{(h)}$,
by integration: $Y^{(h)}(x) = -{\d F^{(h)} / \d V_1(x)}$.

\end{itemize}

\newsection{Leading order: algebraic equation}
\label{sectiontheloopequation}

While we study the leading order, we will drop the subscripts $(0)$ for lisibility.

\medskip

\subsection{Reminder: one-cut asumption for the 1-matrix model}
\label{cutgenusasumption}

In the 1-matrix model, the existence of the $1/N^2$ topological expansion
is related to the 1-cut asumption.
We refer the reader to \cite{BDE, dkmvz, EML} for more detailed explanations,
 and we just summarize the main idea.

When there is more than one cut, i.e. when the density of eigenvalues has a disconnected support,
there are tunnel effects between the different cuts which generate an oscillating function of $N$ (a theta-function).
Such a periodic function cannot have a $1/N^2$ expansion.
The frequencies of oscillations are the filling fraction of eigenvalues in the cuts,
in other words, the frequencies are equal to the integrals of $W(x)$ along the non-trivial cycles of the hyper-elliptical surface generated by $(x,W(x))$
(In the 1-matrix case the surface is hyperelliptical, i.e. $W(x)$ obeys a quadratic equation).
It is only in the one-cut case that there is no oscillations,
and the $1/N^2$ expansion can exist.

\medskip

We expect the same idea to hold for the 2-matrix model.
The method of \cite{BDE} can be adapted to the 2-matrix case:
the frequencies are the contour integrals of $W(x)dx$,
which are the same as the contour integrals of $Y(x)dx$
(which are also the same as the integrals of $X(y)dy$ by integration by parts),
along the non-trivial cycles of the algebraic curve $(x,Y(x))$.
The $1/N^2$ expansion cannot exist if there are oscillations.
Therefore we will require that there is no oscillation,
 i.e. that there is no non-trivial cycle,
i.e. that the algebraic curve has genus zero.

\medskip

Notice a difference between the 1-matrix case and the 2-matrix case:\par
The number of connected parts of the support of the density
is actually the number of cuts in the first sheet only.
There may be cuts in other sheets. \par
- In the 1-matrix case, the curve is hyper-elliptical,
it has only two sheets,
thus the first sheet contains all the cuts.
The number of non-trivial cycles is equal to the number of cuts minus one.
Therefore in the 1-matrix model,
 the genus zero asumption is equivalent to the 1-cut asumption and is equivalent to a density with connected support.\par
- In the 2-matrix case, the curve is only algebraic.
It can have more than two sheets, so that the non-trivial cycles are not necessarily in the first sheet.
One cut in the first sheet is not equivalent to genus zero.
In fact, there are $d_2+1$ sheets and at least $d_2$ cuts,
 $d_2-1$ of them are NOT in the first sheet.
In the 2-matrix case, the connectedness of the support of the density is a necessary condition, but is not sufficient.
The necessary and sufficient condition is the genus zero condition.\par

\medskip

Therefore, in the 2-matrix model,
 we claim that the existence of the $1/N^2$ expansion
 is related to the asumption that the algebraic curve $E^{(0)}(x,y)=0$ has genus zero.

From now on, we will consider only the genus zero case.
We would like to emphasize that many of the results derived below for the leading order loop functions,
 can be derived for arbitrary genus with no difficulty (for instance the functions $\Omega$ are always the Bergmann kernels),
but are out of the scope of this article.
Many of them are presented in \cite{BEHansatz}.

\bigskip
The previous discussion applies to the case where the partition function \eq{Zdef} is well defined,
i.e. the potentials satisfy condition \eq{Zexistencecondition}.
However, there is another case of ``existence'' of the $1/N^2$ expansion:
when the genus $h$ free energy $F^{(h)}$ is formally defined by its diagrammatic expansion \eq{Fhsurfaces} and \eq{Fsurfaces}.
The formal series for $F^{(h)}$ is obtained by expanding in the vicinity of a gaussian integral \cite{ZJDFG, BIPZ},
i.e. by considering that the potentials are ``small deformations'' of quadratic potentials.
When both potentials are quadratic, the algebraic curve $E(x,y)=0$ clearly has genus zero.
The formal expansion in power series of the $g_k$'s and $g^*_l$'s can not change the genus.
Therefore, the formal diagrammatic expansion of $F^{(h)}$ is also obtained by a genus zero asumption.

\subsection{Genus zero asumption}

Since we require that $E(x,y)=0$ be a genus zero algebraic curve,
 there must exist a rational uniformization.
Such a rational uniformization appears naturally in the framework of bi-orthogonal polynomials as shown in \cite{BEHneeds,eynard, eynardchain,DKK}.
That leads us to choose it of the following form:

\beq
x=\xx(s) = \gamma s + \sum_{k=0}^{d_2} {\alpha_k\over s^k}
\virg
y=\yy(s) = {\gamma\over s} + \sum_{j=0}^{d_1} \beta_j s^j
\eeq

This means that for every $(x,y)$ which satisfy $E(x,y)=0$,
 there exists at least one $s$ such that $x=\xx(s)$ and $y=\yy(s)$.

\medskip

We note the inverse functions:
\beq
x=\xx(s) \leftrightarrow s=\ss(x)
\virg
y=\yy(s) \leftrightarrow s=\tt(y)
\eeq
The functions $\ss(x)$ and $\tt(y)$ are multivalued, we will discuss their sheet structure below.
The functions $Y(x)$ and $X(y)$ are:
\beq
Y(x) = \yy(\ss(x)) \virg X(y) = \xx(\tt(y))
\eeq
They are multivalued too, and their sheet structure will be discussed below.

\subsection{The parameters $\alpha_k$ and $\beta_j$}

The $\alpha$'s and $\beta$'s are merely a reparametrization of the coefficients of the potentials.

Indeed,
since for large $x$ we have: $V'_1(x)-Y(x)=W(x) \sim {1\over x} + O(1/x^2)$ (this corresponds to $s\to\infty$),
 and for large $y$ we have: $V'_2(y)-X(y)= \td{W}(y)\sim {1\over y} + O(1/y^2)$ ($s\to 0$),
we must have:
\bea\label{eqmotion}
\yy(s) - V'_1(\xx(s)) & \mathop\sim_{s\to \infty} & -{1\over \gamma s} + O(1/s^2) \cr
\xx(s) - V'_2(\yy(s)) & \mathop\sim_{s\to 0} & -{s\over \gamma} + O(s^2)
\eea
Both these two equations imply that (compute ${\rm Res} \yy(s)\xx'(s)\D{s} = -{\rm Res} \xx(s)\yy'(s)\D{s}$):
\beq
\encadremath{
\gamma^2 = -1+ \sum_{k=1}^{{\rm min}(d_1,d_2)} \, k \alpha_k \beta_k
}\eeq

And eqs.(\ref{eqmotion}) allow to write the coupling constants $g_l$ and $g^*_l$ as functions of the $\alpha_k$'s and $\beta_k$'s (compute ${\rm Res} V'_1(x) x^{l-1} \D{x}$ and ${\rm Res} V'_2(y) y^{l-1} \D{y}$):
\beq
g_{l+1} = -{1\over \gamma^{l+1}} \sum_{k=-1}^{d_1-l-1} \left( \prod_{p=0}^{d_2} \sum_{i_p=0}^{[d_1/(p+1)]} \right) k{(l+\sum_p i_p)! \over l! \prod_p i_p !} \,\, \alpha_k \beta_{(l+k+1+\sum_{p} (p+1) i_p)} \prod_{p=0}^{d_2} \left( -{\alpha_p\over \gamma}\right)^{i_p}
\eeq
\beq
g^*_{l+1} = -{1\over \gamma^{l+1}} \sum_{k=-1}^{d_2-l-1} \left( \prod_{p=0}^{d_1} \sum_{i_p=0}^{[d_2/(p+1)]} \right) k{ (l+\sum_p i_p)! \over l! \prod_p i_p !} \,\, \beta_k \alpha_{(l+k+1+\sum_{p} (p+1) i_p)} \prod_{p=0}^{d_1} \left( -{\beta_p\over \gamma}\right)^{i_p}
\eeq
where we take the convention that $\beta_j=0$ if $j>d_1$ and $\alpha_j=0$ if $j>d_2$, and $\alpha_{-1}=\beta_{-1}=\gamma$.

As an example, for $l=d_1$ and $l=d_1-1$, the formula reduces to:
\beq
g_{d_1+1} = {\beta_{d_1}\over \gamma^{d_1}}
\virg
g_{d_1} = {\beta_{d_1-1}\over \gamma^{d_1-1}} - d_1 \alpha_0 {\beta_{d_1}\over \gamma^{d_1}}
\virg \dots
\eeq

In principle, it should be possible to revert these formula, and compute the $\alpha$'s and $\beta$'s as functions of the coupling constants.
This can be done at least numerically.\par

{\noindent \em Remark:}\par
Notice that there might exist more than one solution.
Only the solution which leads to an absolute minimum of the leading order free energy $F^{(0)}$ should be considered.
If the potentials $V_1$ and $V_2$ are real (which is the case for most physical models), the density of eigenvalues of the first (resp. second) matrix is the imaginary part of $Y(x)$ (resp. $X(y)$) along the cut,
and the density must be positive.
Any solution for the $\alpha$'s and $\beta$'s which does
not satisfy that is unacceptable.
If there is no acceptable solution, this means that our choice of $V_1$ and $V_2$ does not correspond to a genus zero case.
In general, if $V_1$ and $V_2$ have only one well, or if they are close to quadratic potentials, we have a genus zero solution.

\subsection{Sheet structure of the function $Y(x)$}
\label{sectionsheetstructure}

The function $\ss(x)$ is multivalued, indeed the equation
\beq
\xx(s) = x
\eeq
has $d_2+1$ solutions, which we note
\beq
\{\ss_0(x),\ss_1(x),\dots ,\ss_{d_2}(x)\}
\eeq
and therefore, the function $Y(x)=\yy(\ss(x))$ is multivalued with $d_2+1$ values which we note:
\beq
\{Y_0(x),Y_1(x),\dots ,Y_{d_2}(x)\} \quad {\rm where}\,\, Y_k(x) = \yy(\ss_k(x))
\eeq
We define $\ss(x)=\ss_0(x)$ to be the "physical sheet" solution,
 i.e. the one such that $\sigma(x)\to\infty$ when $x\to\infty$,
i.e. the one for which $W(x)=V'_1(x)-Y(x) \sim O(1/x)$.
The other solutions $\ss_k(x)$ with $k=1,\dots, d_2$ all tend to zero when $x\to\infty$.

\subsubsection{Endpoints and cuts}
\label{sectionendpoints}

The functions $Y(x)$ and $\ss(x)$  have cuts, whith a generic square root behaviour near the endpoints, which means that the derivative with respect to $x$ diverges at the endpoints.
Since $Y'(x)={\yy'(s)\over \xx'(s)}$, the endpoints are solutions of $s=0$ or solutions of the equation:
\beq
\xx'(s)=0  
\eeq
which has $d_2+1$ solutions. We note them:
\beq\label{defek}
\{ e_0, e_1,  \dots , e_{d_2} \}
\eeq
the endpoints are thus located at the $\xx(e_k)$'s, as well as at infinity.
In the physical sheet, the function $Y(x)$ is analytical at infinity, and the only cut in the physical sheet is a line joining two endpoints of the type $\xx(e_k)$.
In the other sheets, the function $Y(x)$ behaves as $x^{1/d_2}$ near infinity, the cuts are lines joining some $\xx(e_k)$ to $\infty$.
There are exactly $d_2$ cuts, in agreement with the fact that the genus is zero.

\medskip

Similarly, the function $X(y)$ has exactly $d_1$ cuts,
 whose endpoints are $\yy(\td{e}_j)$'s and $\infty$,
where
\beq\label{deftdek}
\{ \td{e}_0, \td{e}_1,  \dots , \td{e}_{d_1} \}
\eeq
are the $d_1+1$ roots of
\beq
\yy'(s)=0 \,\, .
\eeq
The cut in the physical sheet is a line joining two endpoints of the type $\yy(\td{e}_k)$,
and the cuts in the other sheets are lines joining some $\yy(\td{e}_k)$ to $\infty$.

\bigskip
As said in the introduction, we assume that the potentials are not critical,
which means that the algebraic curve has no singular points (other than $s=0$ and $s=\infty$).
That means that all the $e_k$ and $\td{e}_l$ must be distinct
(examples of critical potentials are given in \cite{DKK}: those which produce the rational minimal conformal models $(p,q)$).
The non-criticality asumption, is crucial for the derivations we are going to present below: we will often have $\xx''(e_k)$ or $\yy'(e_k)$ in the denominators.

\subsubsection{Sheets in the $s$-plane}

We define the $x$-sheets (resp. $y$-sheets) in the $s$-plane, as domains which contain only one $\ss_k(x)$ (resp. $\td{\ss}_k(y)$) when $x$ (resp. $y$) sweeps the complex plane.
In other words, the function $\xx(s)$ (resp. $\yy(s)$) restricted to one domain is one to one.

The domains are separated by closed curves in the $s$-plane,
 which are the projections of the cuts in the $x-plane$ (resp. $y$-plane).
There is some arbitrariness in the choice of domains and cuts (one could take the cuts as straight lines or any other curves which link the endpoints).
The only constraint is that all the cuts go through the endpoints.

The $x$-sheets (resp. $y$-sheets) have the following properties:\\
- The physical sheet is the sheet which contains $\infty$ (resp. $0$).\\
- All the $d_2$ (resp. $d_1$) other sheets contain $0$ (resp. $\infty$).

And the cuts:\\
- The cut in the physical sheet corresponds to a contour in the $s$-plane which encircles $0$.
If the potentials are real, we expect the cut in the $x$-plane (reps. $y$-plane) to be on the real axis, and its $s$-plane image to be a contour which crosses the real axis, and symmetric with respect to the real axis.
The two endpoints are solutions of $\xx'(s)=0$ (resp. $\yy'(s)=0$).\\
- The other $d_2-1$ (reps. $d_1-1$) cuts are contours in the other sheets.
They all meet in $0$ (resp. $\infty$).
One of their endpoint is a solution of $\xx'(s)=0$ (resp. $\yy(s)=0$) and the other endpoint is at $s=0$ (reps. $s=\infty$).

\bigskip

{\bf notational remark:}
when we write $\ss(x)$ this means that $\ss(x)$ is considered as a multivalued function.
For instance the multivalued function $Y(x)$ is equal to
\beq
Y(x) = \yy(\ss(x))
\eeq
We write $\ss_k(x)$ or $Y_k(x)$ only when we want to specify in which sheet is $x$.
By default, we will consider that $x$ is in the physical sheet, and therefore $\ss_0(x)$ is in the domain which contains $s=\infty$, and $\ss_k(x)$ with $k\geq 1$ is in a domain which contains $s=0$.
Note that any symmetric function of $(\ss_0(x),\dots, \ss_{d_2}(x))$ is a regular function of $x$, i.e. it has no cut.

\subsection{The polynomial $P(x,y)$}

The polynomial $P(x,y)$ is found from $E(\xx(s),\yy(s))=0$, i.e. :
\beq
P(\xx(s),\yy(s)) = (V'_1(\xx(s))-\yy(s))(V'_2(\yy(s))-\xx(s)) +1
\eeq
both sides are Laurent polynomials of $s$ of the same degree.
By indentifying the coefficients of powers of $s$ on both sides, one can compute explicitely all the coefficients of the polynomial $P(x,y)$.
The expression of $P(x,y)$ in terms of $\alpha_k$ and $\beta_j$ can be written explicitely with the determinant formula:
\beq\label{Exydiscr}
E(x,y) = {1\over \gamma^{d_1+d_2}} \, \det\pmatrix{
\gamma & \alpha_0-x & \alpha_1 & \dots & \alpha_{d_2} & 0 & \dots & 0 \cr
0      & \ddots     & &  & & \ddots  & & 0 \cr
\vdots &      & \ddots      &  &   & &\ddots  & \vdots \cr
0 & \dots & 0 & \gamma & \alpha_0-x & \alpha_1 & \dots & \alpha_{d_2}  \cr
\beta_{d_1} & \dots & \beta_1 & \beta_0-y & \gamma &  0 & \dots & 0 \cr
0      & \ddots     & &  & & \ddots  & & \vdots \cr
\vdots &      & \ddots      &  &   & &\ddots  & 0 \cr
0 & \dots &  0 & \beta_{d_1} & \dots & \beta_1 & \beta_0-y & \gamma  \cr
}
\eeq
This determinant of size $d_1+d_2+2$ is the discriminant which vanishes if and only if the Laurent polynomials in $s$: $\xx(s)-x$ and $\yy(s)-y$ have a common root.

There is also the formula:
\beq
E(x,y) = -(g_{d_1+1})^{d_2+1} \, {1\over \gamma^{2d_2}} \,\prod_{i=0}^{d_2} \prod_{j=0}^{d_1} (\gamma \ss_i(x)- \gamma \tt_j(y))
\eeq
indeed, it is a symmetric function of the $\ss_i(x)$ and $\tt_j(y)$,
 with no poles, therefore it is a polynomial in $x$ and $y$,
 and it vanishes when $\xx(s)-x$ and $\yy(s)-y$ have a common root.\par

\medskip

{\noindent \em Remark:}
Had we not made the genus zero asumption, we could determine $P(x,y)$ by a determinant formula\footnote{Thanks to J. Hurtubise for that remark.} very similar to \eq{Exydiscr}.
We can also determine $P(x,y)$ by requiring that $E(x,y)=0$ is a genus $g$ curve, and that all the b-cycle integrals of $Y(x)\D{x}$ vanish.
There is one (or more) solution for all $g$.
For any $g$ between $0$ and $\left[{d_1 d_2-1\over 2}\right]$ we can compute all the loop functions (to leading order), and therefore we can compute the free energy $F^{(0)}(g)$ as a function of the genus $g$.
The actual value of $g$ is the one which minimizes $F^{(0)}$.
The genus zero asumption is thus that the minimum is obtained for $g=0$.\par

\newsection{Leading order loop functions}
\label{sectionleadingorder}

\subsection{Some leading order Loop equations}

The change of variables (we drop the $O(1/N^2)$ terms):
\begin{itemize}
\item
$M_1\to M_1 + \epsilon{1\over x-M_1}{1\over y-M_2}+ \epsilon{1\over y-M_2}{1\over x-M_1}$ implies:
\beq
(y-Y(x))W(x,y) = W(x) - \tilde{U}(x,y) = V'_1(x) - Y(x) - \tilde{U}(x,y)
\eeq

\item $M_2\to M_2 + \epsilon{1\over x-M_1}{1\over y-M_2}+ \epsilon{1\over y-M_2}{1\over x-M_1}$ implies:
\beq\label{eqmotionWxy}
(x-X(y))W(x,y) = \tilde{W}(y) - U(x,y) = V'_2(y) - X(y) - U(x,y)
\eeq

\item
$M_1\to M_1 + \epsilon{1\over x-M_1}{V'_2(y)-V'_2(M_2)\over y-M_2}+ \epsilon{V'_2(y)-V'_2(M_2)\over y-M_2}{1\over x-M_1}$ implies:
\beq\label{eqmotionUxy}
(y-Y(x)) U(x,y) = -P(x,y) + V'_2(y) W(x) - {1\over N}<\tr {1\over x-M_1} V'_2(M_2)>
\eeq

\item
$M_2\to M_2 + \epsilon{V'_1(x)-V'_1(M_1)\over x-M_1}{1\over y-M_2}+ \epsilon{1\over y-M_2}{V'_1(x)-V'_1(M_1)\over x-M_1}$ implies:
\beq
(x-X(y)) \tilde{U}(x,y) = -P(x,y) + V'_1(x) \tilde{W}(y) - {1\over N}<\tr {1\over y-M_2} V'_1(M_1)>
\eeq

\item
 $M_2\to M_2 + \epsilon{1\over x-M_1}$ implies:
\beq\label{eqmotionxV2}
{1\over N}<\tr {1\over x-M_1} V'_2(M_2)> = x W(x) - 1
\eeq

\item
 $M_1\to M_1 + \epsilon{1\over y-M_2}$ implies:
\beq\label{eqmotionyV1}
{1\over N}<\tr {1\over y-M_2} V'_1(M_1)> = y \tilde{W}(y) - 1
\eeq

\end{itemize}

\subsection{The function $U(x,y)$}

Using \eq{eqmotionUxy} and \eq{eqmotionxV2}, we find that to leading order, the function $U(x,y)$ is simply:
\beq\label{Uleading}
\encadremath{
U(x,y) = V'_2(y)-x + {E(x,y)\over y-Y(x)}
}\eeq
Notice that this function has no pole,
 and it is indeed a polynomial in $y$.

\subsection{$W(x,y)$}

Using \eq{eqmotionWxy} and \eq{Uleading}, we find that, to leading order, the function $W(x,y)$ is:
\beq\label{WWleading}
\encadremath{
W(x,y) = 1- {E(x,y)\over (x-X(y))(y-Y(x))}
}\eeq
It can also be written:
\beq
W(x,y) = 1 - \left({\beta_{d_1}\over \gamma}\right)^{d_2}\,  \, {\ss_0(x) \tt_0(y)^{d_2} \over \ss_0(x) - \tt_0(y)} \, \prod_{i=1}^{d_2} \prod_{j=1}^{d_1} (\ss_i(x)-\tt_j(y))
\eeq
it has a pole when $\sigma_0(x)=\td\sigma_0(y)$.

We are now going to compute the two-loop functions.

\subsection{The functions $\Omega(x;x')$ and $\td\Omega(y;x')$}

These functions were first computed by \cite{DKK} using the orthogonal polynomial's technics, they can also be found by another method, namely: they must be rational functions of $\ss(x)$ (resp. $\tt(y)$) and $\ss(x')$, and they are determined completely by their poles and behaviours near $\infty$.
The result is the following:
\bea
\Omega(x;x') & = & {\d \over \d V_1(x')} W(x)= \d_x \d_{x'} \ln{\ss(x)-\ss(x')\over x-x'}   = -{1\over (x-x')^2} + {\ss'(x) \ss'(x')\over (\ss(x)-\ss(x'))^2} \cr
& = & -{1\over (x-x')^2} + {1\over \xx'(\ss(x))\xx'(\ss(x'))(\ss(x)-\ss(x'))^2}
\eea
and since $Y(x)=V'_1(x)-W(x)$ we have:
\beq\label{dYdV}
{\d \over \d V_1(x')} Y(x) = - {\ss'(x) \ss'(x')\over (\ss(x)-\ss(x'))^2} = -\d_x \d_{x'} \ln{(\ss(x)-\ss(x'))}
\eeq
i.e. ${\d \over \d V_1(x')} Y(x)$ is the Bergmann kernel\footnote{
Without the genus zero asumption, one has ${\d \over \d V_1(x')} Y^{(0)}(x)=$ Bergman kernel, but ${\d \over \d V_1(x')} Y(x)=$ square of the Szego kernel $+ O(1/N)$.}.

At $x=x'$ we get the Schwarzian derivative of $\ss(x)$:
\beq
\Omega(x;x) = {1\over 6} {\ss'''(x)\over \ss'(x)} - {1\over 4} {\ss''(x)^2\over \ss'(x)^2}
\eeq

\bigskip

Similarly \cite{DKK}, the function $\td\Omega(y;x')$ is:
\beq
\td\Omega(y;x') = {\d\over \d V_1(x')} \td{W}(y)= - \d{x'} \d_y \ln{(\ss(x')-\tt(y))} = -{\ss'(x')\tt'(y)\over (\ss(x')-\tt(y))^2}
\eeq
i.e.:
\beq
{\d \over \d V_1(x')} X(y) = {\tt'(y) \ss'(x')\over (\tt(y)-\ss(x'))^2} = \d_y \d_{x'} \ln{(\tt(y)-\ss(x'))}
\eeq

\subsection{More loop equations}

\begin{itemize}

\item The change of variables $\delta M_2 = {1\over x-M_1} \tr {1\over x'-M_1}$ implies:
\beq
\left< \tr {1\over x-M_1} V'_2(M_2) \tr {1\over x'-M_1} \right>_{\rm c} = x \Omega(x;x')
\eeq

\item Using this result, the change of variables $\delta M_1 = {1\over x-M_1}{V'_2(y)-V'_2(M_2)\over y-M_2} \tr {1\over x'-M_1} + {\rm h.c.} $ implies:
\bea
(y-Y(x)) U(x,y;x') & = & (V'_2(y)-x-U(x,y))\Omega(x;x') - P(x,y;x') 
\cr
&& - \d_{x'} {U(x,y)-U(x',y)\over x-x'}
\eea
i.e., using \eq{Uleading}:
\beq
(y-Y(x)) U(x,y;x') =  - \Omega(x;x'){E(x,y)\over y-Y(x)} - P(x,y;x') - \d_{x'} {U(x,y)-U(x',y)\over x-x'}
\eeq
which can also be written, using \eq{Uleading} again:
\bea\label{UxyxYloopeq}
(y-Y(x)) U(x,y;x') & = &  - (\Omega(x;x')+{1\over (x-x')^2}){E(x,y)\over y-Y(x)} - P(x,y;x') \cr
&& + \d_{x'} {E(x',y)\over (x-x')(y-Y(x'))}
\eea

\item The change of variables $\delta M_1 = {1\over y-M_2} \tr {1\over x'-M_1}$ implies:
\beq
\left< \tr V'_1(M_1) {1\over y-M_2} \tr {1\over x'-M_1} \right>_{\rm c} = y \tilde\Omega(y;x') - \d_{x'} W(x',y)
\eeq

\item Using this result, the change of variables $\delta M_2 = {V'_1(x)-V'_1(M_1)\over x-M_1}{1\over y-M_2} \tr {1\over x'-M_1} + {\rm h.c.} $ implies:
\beq
(x-X(y))\tilde{U}(x,y;x') = \tilde\Omega(y;x') (V'_1(x)-\tilde{U}(x,y)-y) - P(x,y;x') + \d_{x'} W(x',y)
\eeq
which can also be written:
\beq\label{loopeqtdUxyx}
(x-X(y)) \td{U}(x,y;x') =  - P(x,y;x') - \td\Omega(y;x') {E(x,y)\over x-X(y)} + \d_{x'} W(x',y)
\eeq

\end{itemize}

\subsection{Determination of the Polynomial $P(x,y;x')$}

In particular, if for any $k=0,...,d_2$, we choose $y=Y_k(x)$ in \eq{loopeqtdUxyx}
, we have $X(y)=x$, and using \eq{WWleading}:
\beq
P(x,Y_k(x);x')  = \d_{x'} \left[ {E(x',Y_k(x))\over (x-x')(Y_k(x)-Y(x'))} \right] - \td\Omega(Y_k(x);x') E_x(x,Y_k(x))
\eeq
i.e.
\beq
P(x,Y_k(x);x')  = \d_{x'} \left[ {E(x',Y_k(x))\over (x-x')(Y_k(x)-Y(x'))} + { E_x(x,Y_k(x))\over \yy'(\sigma_k(x))(\sigma_k(x)-\sigma(x'))}  \right]
\eeq
We know that $P$ is a polynomial in $y$ of degree $d_2-1$,
and we know its value in $d_2+1$ points, therefore we can determine it with the interpolation formula (see appendix A),
which, in this case, reduces to:
\beq\label{Pxyx0d2}
P(x,y;x') = \sum_{k=0}^{d_2} {P(x,Y_k(x);x')\over (y-Y_k(x))} \, {E(x,y)\over E_y(x,Y_k(x))}
\eeq
The latter expression seems to be a polynomial of degree $d_2$ in $y$,
 but one can check that the leading term vanishes (see appendix B).
Thus we may also write:
\beq\label{Pxyx1d2}
P(x,y;x') = \sum_{k=1}^{d_2} {P(x,Y_k(x);x')(Y_k(x)-Y_0(x))\over (y-Y_k(x))(y-Y_0(x))} \, {E(x,y)\over E_y(x,Y_k(x))}
\eeq
Both expressions give either:
\bea
& P(x,y;x')  = & \!\!\!  \d_{x'}  \sum_{k=1}^{d_2}  { E_x(x,Y_k(x))E(x,y)(Y_k(x)-Y_0(x))\over \yy'(\sigma_k(x))(\sigma_k(x)-\sigma(x')) E_y(x,Y_k(x))(y-Y_k(x))(y-Y_0(x))} \cr
&& + {E(x',Y_k(x))\over (x-x')(Y_k(x)-Y(x'))}{E(x,y)\over E_y(x,Y_k(x))} {Y_k(x)-Y_0(x)\over (y-Y_k(x))(y-Y_0(x))} \cr
\eea
Using  $E(\xx(s),\yy(s))=0$ we have 
$\yy'(\ss_k(x)) E_y(x,Y_k(x)) = - \xx'(\ss_k(x)) E_x(x,Y_k(x))$, and we can write:
\bea\label{Pkneqzero}
& P(x,y;x')  = & \d_{x'}  \sum_{k=1}^{d_2}  { E(x,y)\over \xx'(\sigma_k(x))(\sigma_k(x)-\sigma(x'))} {Y_0(x)-Y_k(x)\over (y-Y_k(x))(y-Y_0(x))} \cr
&& + {E(x',Y_k(x))\over (x-x')(Y_k(x)-Y(x'))}{E(x,y)\over E_y(x,Y_k(x))} {Y_k(x)-Y_0(x)\over (y-Y_k(x))(y-Y_0(x))} \cr
\eea
or, using \eq{Pxyx0d2}:
\bea\label{Pkincludingzero}
- P(x,y;x') & =  \d_{x'} \sum_{k=0}^{d_2} & { E(x,y)\over \xx'(\sigma_k(x))(\sigma_k(x)-\sigma(x'))} {1\over (y-Y_k(x))} \cr
&& - {E(x',Y_k(x))\over (x-x')(Y_k(x)-Y(x'))}{E(x,y)\over E_y(x,Y_k(x))} {1\over (y-Y_k(x))} \cr
\eea

Eq.(\ref{Pkincludingzero}) is clearly a polynomial in $y$, let us explain how one can check that it is also a polynomial in $x$:
it is symmetric in the $\sigma_k(x)$'s,
 therefore it has no cut (it takes the same value in all sheets),
it must be a rational function of $x$.
The only possible poles could be at $x=x'$, $\xx'(\ss_k(x))=0$ or $E_y(x,Y_k(x))=0$.\par
- Take $x=x'$, more precisely, for some $k\in [0,d_2]$, $\sigma_k(x)=\sigma(x')$,
because of the symmetry one may assume that $k=0$,
 and one sees that expression \eq{Pkneqzero} has no pole at $\sigma_0(x)=\sigma(x')$.\par
- Now choose some $k$ and $x$ such that $\xx'(\ss_k(x))=\epsilon$ is very small.
There must exist some (unique because our potentials are non critical) $j\neq k$ such that $\ss_j(x)-\ss_k(x) = O(\epsilon)$,
and one has $\xx'(\ss_j(x))=-\xx'(\ss_k(x)) + O(\epsilon^2)$.
The poles for $j$ and $k$ thus cancel each other.\par
- Now choose some $k$ and $x$ such that $E_y(x,Y_k(x))=\epsilon$ is very small.
This means there must exist some (unique because our potentials are non critical) $j\neq k$ such that $Y_j(x)-Y_k(x) = O(\epsilon)$,
 and one has $E_y(x,Y_j(x))=-E_y(x,Y_k(x))+O(\epsilon^2)$.
The poles for $j$ and $k$ thus cancel each other in \eq{Pkincludingzero}.

\subsection{Determination of $U(x,y;x')$}

We can now compute $U(x,y;x')$ using \eq{UxyxYloopeq}:
\bea
(y-Y_0(x))U(x,y;x') & =  \d_{x'} &
\sum_{k=1}^{d_2} {E(x,y)\over \xx'(\sigma_k(x))(\sigma_k(x)-\sigma(x'))(y-Y_k(x))} \cr
&& - {E(x,y)\over (x-x')}\sum_{k=0}^{d_2} {E(x',Y_k(x))\over (Y_k(x)-Y(x'))E_y(x,Y_k(x))(y-Y_k(x))} \cr
&& + {E(x',y)\over (x-x')(y-Y(x'))}
\eea
which simplifies to:
\bea
(y-Y_0(x))U(x,y;x') & =  \d_{x'} &
\sum_{k=1}^{d_2} {E(x,y)\over \xx'(\sigma_k(x))(\sigma_k(x)-\sigma(x'))(y-Y_k(x))} \cr
&& + {E(x,y)\over (x-x')(y-Y(x'))}\sum_{k=0}^{d_2} {E(x',Y_k(x))\over E_y(x,Y_k(x))(Y_k(x)-y)} \cr
&& + {E(x',y)-E(x,y)\over (x-x')(y-Y(x'))}
\eea
Note the following identity (see appendix B, Lemma \ref{lemma}):
\beq
\sum_{k=0}^{d_2} {\prod_{j=0}^{d_2} (Y_k(x)-Y_j(x'))\over (Y_k(x)-y)\prod_{j\neq k} (Y_k(x)-Y_j(x)) } = 1 - {\prod_j (y-Y_j(x'))\over \prod_j (y-Y_j(x))}
\eeq
which implies:
\beq
\sum_{k=0}^{d_2} {E(x',Y_k(x))\over (Y_k(x)-y)E_y(x,Y_k(x))} = 1 - {E(x',y)\over E(x,y)}
\eeq
and thus:
\beq
\encadremath{
U(x,y;x')  =  \d_{x'}
\sum_{k=1}^{d_2} {E(x,y)\over \xx'(\sigma_k(x))(\sigma_k(x)-\sigma(x'))(y-Y_k(x))(y-Y_0(x))}
}
\eeq
It is easy to check that this expression is a polynomial in $y$ and goes to $0$ when $x\to\infty$ in the physical sheet ($s_0\to\infty$) or when $x'\to\infty$ in the physical sheet.
It has inverse square root singularities near the endpoints where $\xx'(\sigma_k(x))=0$.

\bigskip

{\noindent \bf Computation at $y=Y_0(x)$}

\beq
U(x,Y_0(x);x')  =  \d_{x'}
\sum_{k=1}^{d_2} {E_y(x,Y_0(x))\over \xx'(\sigma_k(x))(\sigma_k(x)-\sigma(x'))(Y_0(x)-Y_k(x))}
\eeq

\bigskip

{\noindent \bf Computation at $x'=x$}

\beq
U(x,Y_0(x);x)  =
\sum_{k=1}^{d_2} {E_y(x,Y_0(x))\over \xx'(\sigma_k(x)) \xx'(\sigma_0(x))(\sigma_k(x)-\sigma_0(x))^2(Y_0(x)-Y_k(x))}
\eeq

We are now equipped to compute the next to leading order functions...

\newsection{Next to leading order}
\label{sectionsubleading}

\subsection{The one-loop function: computation of $Y^{(1)}$}

We have (see \eq{loopeqY1}):
\beq\label{YonePunknown}
Y^{(1)}(x)  =
{P^{(1)}(x,Y_0(x))\over E_y(x,Y_0(x))}
+ \sum_{k=1}^{d_2} {1\over \xx'(\ss_k(x)) \xx'(\ss_0(x))(\ss_k(x)-\ss_0(x))^2(Y_0(x)-Y_k(x))}
\eeq
where $P^{(1)}$ has degrees $(d_1-1,d_2-1)$, and the coefficient of $x^{d_1-1}y^{d_2-1}$ vanishes.
We thus have $d_1 d_2-1$ unknown coefficients to determine.
On the other hand, there are $2d_1 d_2-2$ values of $s$ for which $E_y(\xx(s),\yy(s))=0$ which are not endpoints (see appendix D, \eq{zeroesEyxY}).
If we write that the poles at all these points cancel in equation \ref{YonePunknown}, we can determine $P^{(1)}(x,y)$.
However, we don't need to determine $P^{(1)}(x,y)$, we will follow another method which allows to determine $Y^{(1)}(x)$ directly.

\medskip

Eq.(\ref{YonePunknown}) is a rational function whose only poles
 are the endpoints, with degree up to 5.
Since  \eq{YonePunknown} behaves as $O(s^{-2})$ when $s\to\infty$,
$Y^{(1)}$ has no pole at $\infty$, 
and since \eq{YonePunknown} behaves as $O(s^{1+d_2})$ when $s\to 0$,
$\xx'(s) Y^{(1)}$ has no pole at $0$.

Therefore we may write:
\beq\label{Yoneansatzf}
Y^{(1)}(\xx(s)) = {1\over \xx'(s)} \sum_{k=0}^{d_2} {A_k\over (s-e_k)^4} + {B_k\over (s-e_k)^3} + {C_k\over (s-e_k)^2} + {D_k\over (s-e_k)}
\eeq
The coefficients $A_k$, $B_k$, $C_k$, $D_k$ are determined 
by matching the poles in \eq{YonePunknown} ($P^{(1)}$ doesnot contribute to them).
We will see below that $D_k=0$. 
This is not surprising since we expect $Y^{(1)}(x)$ to be a derivative
 (indeed $Y^{(1)}(x) = -{\d/\d V_1(x)}F^{(1)} = -{d/dx}\,\, {\d/\d V'_1(x)} F^{(1)}$).
 
Let $k\in [0,d_2]$, and choose $s$ close to $e_k$:
\beq
s = e_k+\epsilon
\eeq
there must exist $\td{s}$ (unique because the potentials are non-critical) such that $\xx(\td{s})=\xx(s)$ and $\td{s}$ is close to $e_k$ ($\td{s}$ is the $\ss_k(x)$ in \eq{YonePunknown}):
\beq
\td{s} = e_k - \eta \virg \eta = O(\epsilon)
\eeq
By solving $\xx(s)=\xx(\td{s})$ order by order in $\epsilon$ we get:
\beq
\eta = \l \epsilon 
\virg
\l = 1 + r_k\epsilon+ r_k^2\epsilon^2 + (2r_k^3+t_k) \epsilon^3 + O(\epsilon^4)
\eeq
where
\beq
r_k = {1\over 3} {\xx'''(e_k)\over \xx''(e_k)}
\virg
s_k = {1\over 6} {\xx^{IV}(e_k)\over \xx''(e_k)}
\virg
t_k = {1\over 60} {\xx^{V}(e_k)\over \xx''(e_k)} -r_k s_k
\eeq

From \eq{YonePunknown} we must have:
\beq
{A_k + B_k \epsilon + C_k \epsilon^2 + D_k \epsilon^3}  =   {\epsilon^4\over \xx'(e_k-\eta) (\epsilon+\eta)^2 (\yy(e_k+\epsilon)-\yy(e_k-\eta))} + O(\epsilon^4) 
\eeq
We note the 3rd degree polynomial:
\beq
P_k(\epsilon) = A_k + B_k \epsilon + C_k \epsilon^2 + D_k \epsilon^3
\eeq
i.e.
\bea
P_k(\epsilon)
& = & 
(1+\l)^{-2} \cr
&& (-\l \xx''(e_k) + \epsilon{\l^2\over 2} \xx'''(e_k) - \epsilon^2{\l^3\over 6} \xx^{IV}(e_k) + \epsilon^3{\l^4\over 24} \xx^{V}(e_k))^{-1}  \cr
&& (  (1+\l)\yy'(e_k)
+  {\epsilon\over2}(1-\l^2)\yy''(e_k)
+ {\epsilon^2\over 6}(1+\l^3)\yy'''(e_k) \cr
&& + {\epsilon^3\over 24}(1-\l^4)\yy^{IV}(e_k) )^{-1}  + O(\epsilon^4) \cr
 & = & 
- (\xx''(e_k)\yy'(e_k))^{-1} 
\,\,\, (1+\l)^{-3} \l^{-1} \cr
&& (1- {3\over 2}\epsilon \l r_k 
+ \epsilon^2 \l^2 s_k 
- {5\over 2}\epsilon^3  (t_k+r_k s_k))^{-1}  \cr
&& (  1
+  {\epsilon\over2}(1-\l){\yy''(e_k)\over \yy'(e_k)}
+ {\epsilon^2\over 6}(1-\l+\l^2){\yy'''(e_k)\over \yy'(e_k)} )^{-1} \cr
&&  + O(\epsilon^4)
\eea
It is easy to see that $t_k$ as well as $\yy^{IV}(e_k)$ disappear,
and we find that $D_k=0$.

After a straightforward calculation, one finds:
\bea
&& -8\xx''(e_k) \yy'(e_k) (A_k + B_k \epsilon + C_k \epsilon^2 + D_k\epsilon^3) \cr
&& = 
1-{1\over 3}\epsilon {\xx'''(e_k)\over \xx''(e_k)}
+{1\over 6}\epsilon^2\left({\xx'''(e_k)^2\over \xx''(e_k)^2}
-{\xx^{IV}(e_k)\over \xx''(e_k)}  
+ {\xx'''(e_k)\yy''(e_k)\over \xx''(e_k)\yy'(e_k)}
 - {\yy'''(e_k)\over \yy'(e_k)}\right)  \cr
\eea
After substitution into \eq{Yoneansatzf},
 we find the genus one correction to the resolvent:
\bea
Y^{(1)}(\xx(s))
& = & -\sum_{k=0}^{d_2} {1\over 8 \xx'(s) \xx''(e_k) \yy'(e_k)(s-e_k)^4} \cr
&&  \sum_{k=0}^{d_2} {\xx'''(e_k)\over 24 \xx'(s) \xx''^2(e_k) \yy'(e_k)(s-e_k)^3} \cr
&& - \sum_{k=0}^{d_2} { 
{\xx'''(e_k)^2\over \xx''(e_k)^2}
-  {\xx^{IV}(e_k)\over \xx''(e_k)} 
 + {\xx'''(e_k)\over \xx''(e_k)}{\yy''(e_k)\over \yy'(e_k)} 
- {\yy'''(e_k)\over \yy'(e_k)}
\over 48 \xx'(s) \xx''(e_k) \yy'(e_k)(s-e_k)^2}
 \cr
\eea
That function represents the partition function of a statistical physics model on a genus one surface with one + boundary.

\subsection{The free energy}

We are now going to find the free energy $F^{(1)}$ such that:
\beq
Y^{(1)}(x) = - {\d F^{(1)}\over \d V_1(x)}
\eeq
For that purpose,
 we need to compute the derivatives of various quantities with respect to $V_1(x)$.
From our knowledge of the one matrix case \cite{ACKM},
 we guess that $F^{(1)}$ will be related to the ``moments'' $\yy'(e_k)$.
Therefore, we shall compute the action of ${\d /\d V_1(x)}$ on such quantities.

\subsubsection{The derivatives of $\yy(u)$ and $\xx(u)$}

In this section, we note $x=\xx(s)$ and:
\beq
\dot\xx(u) = \xx'(s){\d \xx(u)\over \d V_1(x)}
\virg
\dot\yy(u) = \xx'(s){\d \yy(u)\over \d V_1(x)}
\eeq

Notice that (using \eq{dYdV} and ${\d \yy(u)\over \d V_1(x)} = {\d Y(\xx(u))\over \d V_1(x)} = {\d \xx(u)\over \d V_1(x)} Y'(\xx(u)) + \left. {\d Y(x')\over \d V_1(x)}\right|_{x'=\xx(u)}$):
\beq\label{dxdVdydV}
\yy'(u) \dot\xx(u) - \xx'(u) \dot\yy(u)  = {1\over (s-u)^2}
\eeq
In particular, if $e_k$, $k=0,\dots,d_2$ is an endpoint $\xx'(e_k)=0$ we have:
\beq
\dot\xx(e_k) = {1\over \yy'(e_k)(s-e_k)^2}
\eeq

Now, notice that  $\dot\xx(u)$ is a Laurent Polynomial\footnote{Indeed $d_2$ is independent of $V_1(x)$. $\dot \yy(u)$ is not a Laurent polynomial, because the degree of $\yy(u)$ depends on $V_1$.} in $u$:
\beq
\dot\xx(u) = \dot\gamma u + \sum_{k=0}^{d_2} \dot\alpha_k u^{-k}
\eeq
where
\beq
\dot\gamma = \xx'(s){\d \gamma\over \d V_1(x)} \virg \dot\alpha_k = \xx'(s){\d \alpha_k\over \d V_1(x)} \quad k=0,\dots, d_2
\eeq
We know its value at all $e_k$'s, $k=0\dots d_2$,
therefore (see appendix A):
\beq
\dot\xx(u) = u\xx'(u) \left[{\dot\gamma\over \gamma}   
+  \sum_{k=0}^{d_2} {1\over e_k (s-e_k)^2 (u-e_k)  \xx''(e_k)  \yy'(e_k) } \right]
\eeq
Since $\alpha_{d_2} = g^*_{d_2+1} \gamma^{d_2}$,
by comparing the coefficient of $u^1$ and $u^{-d_2}$,
we find:
\beq
{\dot\gamma\over \gamma} = 
{1\over d_2} {\dot \alpha_{d_2}\over \alpha_{d_2}} =
 {1\over 2}  \sum_{k=0}^{d_2} {1\over (s-e_k)^2  e_k^{2} \xx''(e_k)  \yy'(e_k) }
\eeq
and thus:
\beq\label{dxdV}
\xx'(s){\d \xx(u)\over \d V_1(x)} 
= \dot\xx(u) 
=  {u\xx'(u)\over 2} \sum_{k=0}^{d_2} {u+e_k\over u-e_k}{ 1\over (s-e_k)^2 e_k^{2} \xx''(e_k)  \yy'(e_k) }
\eeq
Notice that, as a function of $s$,
 this expression has double poles at the endpoints $e_k$.

For any $e_i$, $i=0,\dots,d_2$, $\xx'(e_i)=0$ implies that $\xx'(s){\d e_i\over \d V_1(x)} = - {\dot\xx'(e_i)\over \xx''(e_i)}$
and thus:
\bea\label{dedV}
\dot{e_i} = \xx'(s){\d e_i\over \d V_1(x)} 
= -{\dot\xx'(e_i)\over \xx''(e_i)} 
& = & -{e_i\over 2} \sum_{k\neq i} {e_i+e_k\over e_i-e_k}{ 1\over (s-e_k)^2 e_k^{2} \xx''(e_k)  \yy'(e_k) } \cr
&& - { 3 + e_i{\xx'''(e_i)\over \xx''(e_i)}\over 2 e_i (s-e_i)^2  \xx''(e_i)  \yy'(e_i) }
\eea

From \eq{dxdVdydV} and \eq{dxdV} we have:
\beq
\dot\yy(u) = {1\over \xx'(u)}\left( \yy'(u)\dot\xx(u) - {1\over (s-u)^2} \right)
\eeq
which implies:
\bea
\xx'(s){\d \yy'(e_i)\over \d V_1(x)} = &  \dot\yy'(e_i) + & \dot{e_i} \yy''(e_i) \cr
= & {1\over 2\xx''(e_i)}( &
\dot\xx(e_i) \yy'''(e_i) + \dot\xx''(e_i) \yy'(e_i) - {6\over (s-e_i)^4}   \cr
&& - \dot\xx(e_i) \yy''(e_i) {\xx'''(e_i)\over \xx''(e_i)} - \dot\xx'(e_i) \yy'(e_i) {\xx'''(e_i)\over \xx''(e_i)} \cr
&& + {2\over (s-e_i)^3}{\xx'''(e_i)\over \xx''(e_i)})\cr
\eea
This expression has poles of degree $2$, $3$ and $4$, at the $e_k$'s.
We write:
\beq
\xx'(s){\d \yy'(e_i)\over \d V_1(x)} = \sum_{k=0}^{d_2} {A_{k,i}\over (s-e_k)^4} + {B_{k,i}\over (s-e_k)^3} + {C_{k,i}\over (s-e_k)^2}
\eeq
with:
\beq\label{AkiBki}
A_{k,i} = -3{\delta_{k,i}\over \xx''(e_i)}
\virg
B_{k,i} =  {\delta_{k,i} \xx'''(e_i)\over \xx''(e_i)^2}
\eeq
and:
\beq
C_{k,i} =  
{\left(
 g_k(e_i) (\yy'''(e_i) - \yy''(e_i) {\xx'''(e_i)\over \xx''(e_i)})
 - g_k'(e_i) \yy'(e_i) {\xx'''(e_i)\over \xx''(e_i)} 
 + g_k''(e_i) \yy'(e_i)
\right)
\over 2 \xx''(e_i)e_k^{2} \xx''(e_k)  \yy'(e_k)}
\eeq
where
\beq
g_k(u) =  {u\xx'(u)\over 2}  {u+e_k\over u-e_k}
\eeq
in particular:
\beq
g_i(e_i) = e_i^2 \xx''(e_i)
\virg
g'_i(e_i) = {3\over 2}e_i \xx''(e_i) + {1\over 2} e_i^2 \xx'''(e_i)
\eeq
\beq
g''_i(e_i) = \xx''(e_i) + {3\over 2} e_i \xx'''(e_i) 
+ {1\over 3} e_i^2 \xx^{IV}(e_i)
\eeq
and if $k\neq i$:
\beq
g_k(e_i) = 0
\virg
g'_k(e_i) = {e_i\over 2}{e_i+e_k\over e_i-e_k} \xx''(e_i)
\eeq
\beq
g''_k(e_i) = {e_i+e_k\over e_i-e_k} \xx''(e_i)
 - 2 {e_i e_k\over (e_i-e_k)^2} \xx''(e_i)
 +  {e_i\over 2}{e_i+e_k\over e_i-e_k} \xx'''(e_i)
\eeq

We thus have:
\beq\label{Cii}
C_{i,i}  =   
{ \yy'''(e_i) - \yy''(e_i) {\xx'''(e_i)\over \xx''(e_i)}\over 2 \xx''(e_i)  \yy'(e_i)} 
 - {  \xx'''(e_i)^2 \over 4 \xx''(e_i)^3  }   
+ {1 \over 2 e_i^{2} \xx''(e_i)  }   
+ {\xx^{IV}(e_i) \over 6  \xx''(e_i)^2  } 
\eeq
and if $k\neq i$:
\beq\label{Cki}
C_{k,i}  = 
 {(e_i^2-e_k^2-2e_i e_k) \yy'(e_i) 
\over 2 e_k^{2} (e_i-e_k)^2 \xx''(e_k)  \yy'(e_k)} 
\eeq

Now, let us compute the following quantity (our guess is that it is related to $Y^{(1)}$):
\bea\label{sumdyedV}
\xx'(s)  \sum_{i=0}^{d_2} {\d \ln\yy'(e_i)\over \d V_1(x)} & = & 
\sum_{k=0}^{d_2} {A_{k,k} \over \yy'(e_k)(s-e_k)^4} 
 + \sum_{k=0}^{d_2} {B_{k,k} \over \yy'(e_k)(s-e_k)^3} \cr
&& + \sum_{k=0}^{d_2} {C_{k,k} \over \yy'(e_k)(s-e_k)^2} 
 + \sum_{k=0}^{d_2} \sum_{i\neq k} {C_{k,i} \over \yy'(e_i)(s-e_k)^2} \cr
\eea

substituting eqs.(\ref{AkiBki}, \ref{Cii}, \ref{Cki}) into \eq{sumdyedV}, we get:
\bea
\xx'(s)  \sum_{i=0}^{d_2} {\d \ln\yy'(e_i)\over \d V_1(x)} & = & 
- 3 \sum_{k=0}^{d_2} {1\over \xx''(e_k)\yy'(e_k) (s-e_k)^4} \cr
&& + \sum_{k=0}^{d_2} {\xx'''(e_k)\over \xx''(e_k)^2 \yy'(e_k)(s-e_k)^3} \cr
&& + \sum_{k=0}^{d_2} { 
{\yy'''(e_k)\over \yy'(e_k)} 
- {\yy''(e_k)\over \yy'(e_k)} {\xx'''(e_k)\over \xx''(e_k)} 
-{1\over 2} {\xx'''(e_k)^2\over \xx''(e_k)^2}
+{1\over 3} {\xx^{IV}(e_k)\over \xx''(e_k)}
\over 2 \xx''(e_k) \yy'(e_k) (s-e_k)^2} \cr
&& + \sum_{k=0}^{d_2} 
 {1 \over 2 e_k^{2} \xx''(e_k) \yy'(e_k)(s-e_k)^2} \cr
&& + \sum_{k=0}^{d_2} \sum_{i\neq k} 
 {(e_i^2-e_k^2-2e_i e_k)  
\over 2 e_k^{2} (e_i-e_k)^2 \xx''(e_k)  \yy'(e_k) (s-e_k)^2} \cr
\eea

Notice the following identity\footnote{One proves it by taking the log derivative of $u^{d_2+1}\xx'(u)=\prod_k(u-e_k)$}:
\bea
1+\sum_{i\neq k} {(e_i^2-e_k^2-2e_i e_k)  \over  (e_i-e_k)^2}
 &=&
d_2+1 - 2 e_k^2 \sum_{i\neq k} {1\over  (e_i-e_k)^2} \cr
&=& -(d_2+1) + {2\over 3} e_k^2 {\xx^{IV}(e_k)\over \xx''(e_k)} 
- {e_k^2\over 2} {\xx'''(e_k)^2\over \xx''(e_k)^2} \cr
\eea

therefore:
\bea
\xx'(s) \sum_{i=0}^{d_2} {\d \ln\yy'(e_i)\over \d V_1(x)} & = & 
- 3 \sum_{k=0}^{d_2} {1\over \xx''(e_k)\yy'(e_k) (s-e_k)^4} \cr
&& + \sum_{k=0}^{d_2} {\xx'''(e_k)\over \xx''(e_k)^2 \yy'(e_k)(s-e_k)^3} \cr
&& + \sum_{k=0}^{d_2} { 
{\yy'''(e_k)\over \yy'(e_k)} 
- {\yy''(e_k)\over \yy'(e_k)} {\xx'''(e_k)\over \xx''(e_k)} 
- {\xx'''(e_k)^2\over \xx''(e_k)^2}
+ {\xx^{IV}(e_k)\over \xx''(e_k)}
\over 2 \xx''(e_k) \yy'(e_k) (s-e_k)^2} \cr
&& -(d_2+1) \sum_{k=0}^{d_2} 
 {1 \over 2 e_k^{2} \xx''(e_k) \yy'(e_k)(s-e_k)^2} \cr
\eea

We recognize:
\beq
 \sum_{i=0}^{d_2} {\d \ln\yy'(e_i) \over \d V_1(x)} 
= 24\, Y^{(1)}(x) - (d_2+1) {\d \ln\gamma\over \d V_1(x)} 
\eeq

\subsubsection{The free energy}

The free energy is such that:
\beq
W(x) = {\d F\over \d V_1(x)} = V'_1(x) - Y(x)
\eeq
therefore to order $1/N^2$ we have:
\beq
{\d F^{(1)}\over \d V_1(x)} = - Y^{(1)}(x)
\eeq
which implies (up to a constant $K$ independent of $V_1$):
\beq
F^{(1)} = -{1\over 24} \ln{\left( \gamma^{4} D\right)} + K
\eeq
where:
\beq
D=  {1\over \gamma^{d_1+d_2+2}}
\det\pmatrix{
-\gamma & 0 & \alpha_1 & \dots & d_2\alpha_{d_2} & 0 & \dots & 0 \cr
0      & \ddots     & &  & & \ddots  & & 0 \cr
\vdots &      & \ddots      &  &   & &\ddots  & \vdots \cr
0 & \dots & 0 & -\gamma & 0 & \alpha_1 & \dots & d_2\alpha_{d_2}  \cr
d_1\beta_{d_1} & \dots & \beta_1 & 0 & -\gamma &  0 & \dots & 0 \cr
0      & \ddots     & &  & & \ddots  & & \vdots \cr
\vdots &      & \ddots      &  &   & &\ddots  & 0 \cr
0 & \dots &  0 & d_1\beta_{d_1} & \dots & \beta_1 & 0 & -\gamma  \cr
}
\eeq
$D$ is such that:
\beq
D = (-1)^{d_1+1} \left({d_2\alpha_{d_2}\over \gamma}\right)^{2}
\prod_{\xx'(e)=0} {\yy'(e)\over \gamma}
= (-1)^{d_1+d_2} \gamma^{2}
\prod_{\yy'(\td{e})=0} \left(\xx'(\td{e})\over \gamma\right)
\eeq
The Constant $K$ is independent of $V_1$,
 and if we repeat the calculation by exchanging the roles of $M_1$ and $M_2$,
we find the same expression for $F^{(1)}$ with $K$ independent of $V_2$ as well.
By comparing with the gaussian case we get $K=0$.

We have found the genus one free energy:
\beq\label{Fgenusone}
\encadremath{
F^{(1)} = -{1\over 24} \ln{\left( \gamma^{4} D\right)}
}\eeq

\bigskip

{\noindent \it Remark:}
$D$ is a determinant which vanishes everytime the surface is singular, 
i.e. when the potentials are critical.
We see that near a critical point, the genus one free energy has a logarithmic divergence.
This is in agreement with known double-scaling limit exponents.

\subsection{Higher genus expansion}

The procedure can be repeated to higher orders.
We have the three equations:
\beq\label{loopeqEexact}
E(x,Y(x)) = (V'_1(x)-Y(x))(V'_2(Y(x))-x) - P(x,Y(x)) +1 = {1\over N^2} U(x,Y(x);x)
\eeq
\beq\label{loopeqUexact}
(y-Y(x)) U(x,y) = (V'_1(x)-Y(x))(V'_2(y)-x) - P(x,y) + 1 - {1\over N^2} U(x,y;x)
\eeq
\beq\label{loopinsrelUU}
U(x,y;x') = {\d U(x,y)\over \d V_1(x')}
\eeq
Imagine that we already know the functions $Y(x)$, $E(x,y)$ (i.e. $P(x,y)$), $U(x,y)$ and $U(x,y;x')$
up to order $h-1$.
Let us write \eq{loopeqEexact} to order $h$.
we know all terms but $Y^{(h)}(x)$ and $P^{(h)}(x,y)$.
The factor of the $Y^{(h)}(x)$ term is $E_y^{(0)}(x,Y^{(0)}(x))$,
it vanishes for all $(x,y)$ such that $y=Y^{(0)}(x)$ and $E_y^{(0)}(x,Y^{(0)}(x))$,
i.e. at $2d_1 d_2-2$ points, 
therefore we know $P^{(h)}(x,y)$ in $2d_1 d_2-2$ points.
Since $P^{(h)}(x,y)$ is a polynomial of degree $(d_1-1, d_2-1)$,
 with $d_1 d_2 -1 $ unknown coefficients,
 $P^{(h)}(x,y)$ can be determined completely.
Then, from \eq{loopeqEexact}, $Y^{(h)}(x)$ can be determined.
Using \eq{loopeqUexact}, we can determine $U^{(h)}(x,y)$,
 and using \eq{loopinsrelUU} we can determine $U^{(h)}(x,y;x')$ to order $h$.

We clearly have an algorithmic recursive procedure to find $Y^{(h)}(x)$ to any order $h$.

Finding the free energy $F^{(h)}$ is more difficult,
 as one needs to integrate with respect to the potentials.
The easiest is to have an ansatz for $F^{(h)}$ and check that its derivative with respect to $V_1(x)$ is indeed $-Y^{(h)}(x)$, as in \cite{ACKM}.

Define:
\beq
M_{0,i} = \yy'(e_i) \virg M_{k,i} = {\yy^{(k+1)}(e_i)\over \yy'(e_i)} \,\,\, k\geq 1
\eeq
\beq
J_{0,i} = \xx''(e_i) \virg J_{k,i} = {\xx^{(k+2)}(e_i)\over \xx''(e_i)} \,\,\, k\geq 1
\eeq
We expect the genus $h>1$ free energy to be a rational function of these quantities, of the form:
\beq
F^{(h)} = \sum_{r,s,p,q,m\geq 0} \sum_{k_1,\dots, k_s}\sum_{l_1,\dots, l_s}
 A_{{k_1,\dots, k_s},{l_1,\dots, l_s},p,q,m} {1\over \gamma^{m}} 
\prod_{i=0}^{d_2} {\prod_{j=1}^r \prod_{u=1}^s M_{k_j,i} J_{l_u,i}\over M_{0,i}^{p} J_{0,i}^{q}}
\eeq
with the restrictions that:
\beq
p+q = \sum_k k_j + \sum_u l_u
\eeq
%
The selection rules (of the form $m+p+q \leq 4h-4$ as in \cite{ACKM}) are still to be understood.

\newsection{The one matrix case: $d_2=1$}
\label{section1mat}

In this section we check that for $d_2=1$, our results reduce to those of \cite{ACKM}.

Consider $d_2=1$ and $V_2(y)={1\over 2} g^*_2 y^2 + g^*_1 y$.
$V_2$ is thus quadratic and $M_2$ can easily be integrated out.
The 2-matrix model in this case reduces to a one matrix model with the potential:
\beq\label{VV1}
V(x) = V_1(x) - {1\over 2g^*_2} x^2 + {g^*_1\over g^*_2} x
\eeq

The function $\xx(s)$ and $\yy(s)$ are:
\beq\label{xy1mat}
\xx(s)= \gamma s + \alpha_0 + {\alpha_1\over s}
\virg
\yy(s) ={\gamma\over s} + \sum_{k=0}^{d_1} \beta_k s^k
\eeq
From \eq{eqmotion} we find that:
\beq
\alpha_1 = g^*_2 \gamma
\virg
\alpha_0 = g^*_2 \beta_0 + g^*_1
\eeq
and:
\beq\label{V1yy1mat}
V_1'(\xx(s)) = \yy(s) + \yy(g^*_2/s) - \beta_0 - {\gamma\over s} - {\gamma \over g^*_2} s
\eeq
indeed this expression is a polynomial in $x$ and it satisfies \eq{eqmotion}.
It follows from \eq{VV1}:
\beq\label{Vyy1mat}
V'(\xx(s)) = \yy(s) + \yy(g^*_2/s) - 2(\beta_0 + {\gamma\over s} + {\gamma \over g^*_2} s )
\eeq

The endpoints $a$ and $b$ are the roots of $\xx'(s)$ i.e. they correspond to:
\beq
e=\sqrt{g^*_2} \virg a=\xx(-e) \virg b= \xx(+e)
\eeq
The resolvent (defined in \eq{defWx}) can be written \cite{ACM}:
\beq
W(x) = V'_1(x)-Y(x) = {1\over 2} \left( V'(x) - M(x) \sqrt{(x-a)(x-b)} \right)
\eeq
where $M(x)$ is a polynomial in $x$ (the same as in \cite{ACM}).
Using \eq{Vyy1mat} and \eq{xy1mat} we find that:
\beq
M(\xx(s)) = {\yy(s)-\yy(g^*_2/s)\over s\xx'(s)}
\virg
\sqrt{(\xx(s)-a)(\xx(s)-b)} = s\xx'(s)
\eeq

In particular at the endpoints we have:
\beq
M(a) = {\yy'(-e)\over \gamma} \virg M(b) ={\yy'(+e)\over \gamma}
\eeq

It is clear now, that our result \eq{Fgenusone} coincides with the genus one free energy found by the authors of \cite{ACM}:
\beq\label{fetorusonemat}
F^{(1)} = -{1\over 24} \ln{\left(\left({b-a\over 4}\right)^4M(a)M(b)\right)}
\eeq

\newsection{Conclusion and prospects}
\label{conclusion}

The recursive method presented in this article, 
allows one to compute the resolvents as well as other loop-functions,
to any order in the topological expansion of the 2-matrix model.
We have found the expression of the genus one free energy, 
but an algorithmic method to compute the free energy to higher orders is still incomplete.
For that purpose, it would be usefull to define some ``moments'' 
 and basis functions, as in \cite{ACKM},
onto which one could decompose the loop-functions,
and selection rules to know what set of moments should appear in the free energy at a given order.

It would be interesting also to understand what terms survive in the double scaling limit when the potentials are near a critical point,
and to develop a recursive method to compute directly the free energy and resolvents in the double scaling limit.

As we said in the introduction,
 we have only developped a method to compute the terms in the $1/N^2$ expansion,
but we have not proven the existence of that expansion.
A rigorous proof of that existence can probably be obtained by Riemann-Hilbert technics \cite{BEHRH}.

It would also be usefull to understand what the $1/N^2$ expansion becomes in the non genus-zero case.
It is very likely that it would involve the $\theta$-functions associated to the algebraic curve, as in \cite{BDE}.

\vspace{1cm}
{\noindent} {\it Aknowledgements:} The author would like to thank the CRM, Prof. J. Harnad and Prof. J. Hurtubise for their support when this work was completed.

\setcounter{section}{0}

\appendix{The interpolation formula}
\label{appendixinterpol}

Let $d$ be an integer, and $x_0,\dots, x_d$ be $d+1$ distinct complex numbers.
We note:
\beq
\pi(x) = \prod_{k=0}^d (x-x_k)
\eeq
Let $P_0,\dots,P_d$ be $d+1$ complex numbers.

There is a unique polynomial $P$ of degree $d$ such that:
\beq
P(x_k)=P_k \qquad k=0,\dots, d
\eeq
That polynomial is:
\beq
P(x) = \sum_{k=0}^d P_k {\pi(x)\over (x-x_k)\pi'(x_k)}
\eeq

Any polynomial $P(x)$ of degree $> d$ such that:
\beq
P(x_k)=P_k \qquad k=0,\dots, d
\eeq
can be written:
\beq
P(x) = Q(x)\pi(x) + \sum_{k=0}^d P_k {\pi(x)\over (x-x_k)\pi'(x_k)}
\eeq
where $Q(x)$ is a polynomial of degree $\deg P-d-1$.

\appendix{Proof that \eq{Pxyx0d2} is a polynomial of degree $d_2-1$ }
\label{appendixlemma}

The leading term of \eq{Pxyx0d2} is (we use \eq{Pkincludingzero}):
\beq
\d{x'} {1\over x-x'} \sum_{k=0}^{d_2} \left[ 
{E(x',Y_k(x))\over (Y_k(x)-Y(x'))}{1\over E_y(x,Y_k(x))} - {x-x'\over \xx'(\ss_k(x))(\ss_k(x)-\ss(x'))}
\right]
\eeq
We have (see \eq{formulasumssx}):
\beq
\sum_{k=0}^{d_2}  {1\over \xx'(\ss_k(x))(\ss_k(x)-\ss(x'))}
\eeq
Therefore we have to prove that:
\beq
\sum_{k=0}^{d_2} {E(x',Y_k(x))\over (Y_k(x)-Y(x'))}{1\over E_y(x,Y_k(x))}
= 1
\eeq
Note that:
\beq
\sum_{k=0}^{d_2} {E(x',Y_k(x))\over (Y_k(x)-Y(x'))}{1\over E_y(x,Y_k(x))}
= \sum_{k=0}^{d_2} {\prod_{j\neq 0} (Y_k(x)-Y_j(x'))\over \prod_{j\neq k} (Y_k(x)-Y_j(x))}
\eeq

{\noindent \bf Lemma:}

For any $Y_0,\dots,Y_{d_2}$ and for any $Y'_1,\dots,Y'_{d_2}$ we have:
\beq\label{lemma}
\sum_{k=0}^{d_2} {\prod_{j\neq 0} (Y_k-Y'_j)\over \prod_{j\neq k} (Y_k-Y_j)}
= 1
\eeq

Proof:
\beq
\sum_{k=0}^{d_2} {\prod_{j\neq 0} (Y_k-Y'_j)\over \prod_{j\neq k} (Y_k-Y_j)}
= 
 \sum_{k=0}^{d_2} (-1)^k f(Y_k) {\Delta_{d_2}(Y_0,\dots,Y_{k-1},Y_{k+1},\dots,Y_{d_2})\over \Delta_{d_2+1}(Y_0,\dots,Y_{d_2})}
\eeq
where $f(Y)=\prod_{j\neq 0}(Y-Y'_j)$ is a monic polynomial of degree $d_2$ in $Y$,
and $\Delta_{n}$ is the Vandermonde determinant of $n$ variables.

We recognize a determinant (which is $1$ by linearly combining columns):
\beq
\sum_{k=0}^{d_2} {\prod_{j\neq 0} (Y_k-Y'_j)\over \prod_{j\neq k} (Y_k-Y_j')}
= {\det\pmatrix{1 & Y_0 & Y_0^2 & \dots & Y_0^{d_2-1} & f(Y_0) \cr
1 & Y_1 & Y_1^2 & \dots & Y_1^{d_2-1} & f(Y_1) \cr
\vdots &  &  &  &  & \vdots \cr
1 & Y_0 & Y_{d_2}^2 & \dots & Y_{d_2}^{d_2-1} & f(Y_{d_2}) \cr }
\over \Delta_{d_2+1}(Y_0,\dots,y_{d_2})}
= 1
\eeq

\appendix{Other loop-functions}
\label{appendixH}

Define:
\bea
H(x,y,x',y') & = & {1\over 2N} \left< \tr {1\over x-M_1}{1\over y-M_2}{1\over x'-M_1}{1\over y'-M_2} \right> \cr
&& + {1\over 2N} \left< \tr {1\over y'-M_2}{1\over x'-M_1}{1\over y-M_2}{1\over x-M_1}\right> \cr
&& = {\epsfysize=2.truecm\epsfbox{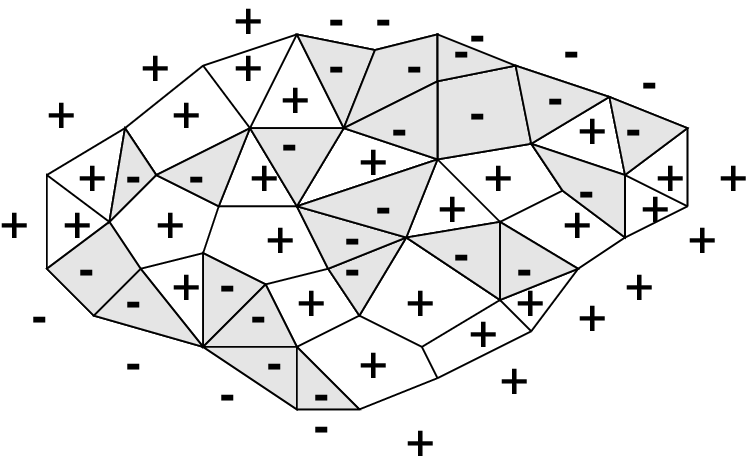}}
\eea
Notice that we have:
\beq
H(x,y,x',y') = H(x',y',x,y) = H(x,y',x',y)
\eeq
We also define:
\bea
F(x,y,x',y') & = & {1\over 2N} \left< \tr {V'_1(x)-V'_1(M_1)\over x-M_1}{1\over y-M_2}{1\over x'-M_1}{1\over y'-M_2} \right> \cr
&& + {1\over 2N} \left< \tr {1\over y'-M_2}{1\over x'-M_1}{1\over y-M_2}{V'_1(x)-V'_1(M_1)\over x-M_1} \right>
\eea
we have:
\beq
F(x,y,x',y') = F(x,y',x',y)
\eeq

{\noindent Loop equations:}\par

The invariance under the change of variable\\
$\delta M_1 = {1\over x-M_1}{1\over y-M_2}{1\over x'-M_1}{1\over y'-M_2} + {\rm h.c.} $ implies:
\beq
(y'-Y(x))H(x,y,x',y') = -{W(x,y)-W(x',y)\over x-x'}(1-W(x',y')) - F(x,y,x',y')
\eeq
The same equation with $y\leftrightarrow y'$ gives:
\bea
&& H(x,y,x',y') =  {1\over (x-x')(y-y')} \left[ {E(x',y)E(x,y')\over (x'-X)(y-Y')(x-X')(y'-Y)} \right. \cr
&& - \left. {E(x,y)E(x',y')\over (x-X)(y-Y)(x'-X')(y'-Y')} \right]
\eea

it follows:
\bea
F(x,y,x',y') & = &
{E(x,y)E(x',y')\over (x-x')(y-y')(x-X)(x'-X')(y'-Y')} \cr
&& - {E(x',y)E(x,y')\over (x-x')(y-y') (x-X')(x'-X)(y-Y')} \cr
&& - {E(x',y)E(x',y')\over (x-x')(x'-X)(x'-X')(y-Y')(y'-Y')}
\eea

\appendix{Some relations}
\label{appendixformula}

The following formula are usefull for many calculations in this article.
They are easy to prove.

\beq
E(x,y) = A \prod_{i,j} (\sigma_i(x)-\td\sigma_j(y))
\eeq
with
\beq
A = (-1)^{d_2+1} \gamma^{1+d_1-d_2} \beta_{d_1}^{d_2} g_{d_1+1} = (-1)^{d_2+1} {\beta_{d_1}^{d_2+1}\over \gamma^{d_2-1}}
\eeq

\beq
\xx(s)-x = {\gamma\over s^{d_2}}\prod_{i=0}^{d_2} (s-\sigma_i(x))
\virg
\yy(s)-y = {\beta_{d_1}\over s}\prod_{j=0}^{d_1} (s-\td\sigma_j(y))
\eeq

\beq
\xx'(\sigma_k(x)) = {\gamma\over \sigma_k^{d_2}(x)} \prod_{i\neq k} (\sigma_k(x)-\sigma_i(x))
\virg
\yy'(\td\sigma_k(y)) = {\beta_{d_1}\over \td\sigma_k(y)} \prod_{j\neq k} (\td\sigma_k(y)-\td\sigma_j(y))
\eeq

\beq\label{formulasumssx}
\sum_{i=0}^{d_2} {1\over \xx'(\sigma_i(x))(s-\sigma_i(x))} = {1\over \xx(s)-x}
\virg
\sum_{j=0}^{d_1} {1\over \yy'(\td\sigma_j(y))(s-\td\sigma_j(y))} = {1\over \yy(s)-y}
\eeq

\bea
E_y(x,Y_0(x)) & = & -A {1\over \yy'(\ss_0(x))} \prod_{(i,j)\neq (0,0)} (\sigma_i(x)-\td\sigma_j(Y_0(x))) \cr
&=& {\beta_{d_1}^{d_2}\over \gamma^{d_2}}  \sigma_0^{d_2+1}(x)
\xx'(\sigma_0(x)) \, \prod_{i\neq 0}\prod_{j\neq 0} (\sigma_i(x)-\td\sigma_j(Y_0(x)))
\eea

\beq
Y_k(x)-Y_0(x) = {\beta_{d_1}\over \sigma_k(x)} \prod_{j} (\sigma_k(x)-\td\sigma_j(Y_0(x)))
\eeq

Limits:
\beq
E_y(\xx(s),\yy(s)) \mathop{\sim}_{s\to \infty} - {\alpha_{d_2}\beta_{d_1}^{d_2}\over \gamma^{d_2}}s^{d_1 d_2}
\virg
E_y(\xx(s),\yy(s)) \mathop{\sim}_{s\to 0} -d_2 {\beta_{d_1}\alpha_{d_2}^{d_1+1}\over \gamma^{d_1+1}} s^{1-d_1 d_2-d_2}
\eeq

$E_y(\xx(s),\yy(s))$ is a Laurent polynomial:
\beq\label{zeroesEyxY}
E_y(\xx(s),\yy(s)) = - {\xx'(s)\over s^{2-d_1 d_2}} P_{2d_1 d_2-2}(s) 
\eeq
where $P_{2d_1 d_2-2}(s)$ is a polynomial of degree $2d_1 d_2 -2$.
Therefore, $E_y(\xx(s),\yy(s))=0$ has $2d_1 d_2 -2$ zeroes.


\end{document}